\documentclass[11pt]{amsart}
\usepackage{amsthm}

\usepackage[all]{xy}
\usepackage{amssymb}
\usepackage{enumerate}
\usepackage{mathrsfs}
\usepackage{epsfig}

\usepackage{graphics}
\usepackage{graphicx}
\usepackage{float}
\usepackage{caption}
\usepackage{subcaption}
\usepackage{epigraph}

\numberwithin{equation}{section}

\newtheorem{thm}{Theorem}[section]

\newtheorem{rem}{Remark}[section]
\newtheorem{example}[thm]{Example}

\newtheorem{defin}[thm]{Definition}

\newcommand{\eq}[1]{(\ref{#1})}

\newcommand{\mbr}{\medbreak}

\renewcommand{\Re}{\operatorname{\rm Re}}
\renewcommand{\Im}{\operatorname{\rm Im}}

\newcommand{\beqast}{\begin{eqnarray*}}
\newcommand{\eqast}{\end{eqnarray*}}
\newcommand{\beqa}{\begin{eqnarray}}
\newcommand{\eqa}{\end{eqnarray}}

\newcommand{\bbe}{\begin{equation}}
\newcommand{\ee}{\end{equation}}

\renewcommand{\Re}{\operatorname{\rm Re}}
\renewcommand{\Im}{\operatorname{\rm Im}}

\newcommand{\bC}{{\mathbb C}}
\newcommand{\bE}{{\mathbb E}}
\newcommand{\bL}{{\mathbb L}}

\newcommand{\bQ}{{\mathbb Q}}
\newcommand{\bX}{{\mathbb X}}

\newcommand{\bR}{{\mathbb R}}

\newcommand{\bZ}{{\mathbb Z}}

\newcommand{\cK}{{\mathcal K}}
\newcommand{\cF}{{\mathcal F}}

\newcommand{\cE}{{\mathcal E}}
\newcommand{\cG}{{\mathcal G}}

\newcommand{\cL}{{\mathcal L}}

\newcommand{\cC}{{\mathcal C}}

\newcommand{\barX}{{\bar X}}
\newcommand{\uX}{{\underline X}}

\newcommand{\cEq}{{\mathcal E_q}}
\newcommand{\cEpq}{{\mathcal E^+_q}}
\newcommand{\cEmq}{{\mathcal E^-_q}}

\newcommand{\phipq}{{\phi^+_q}}
\newcommand{\phimq}{{\phi^-_q}}

\newcommand{\tV}{{\tilde V}}

\newcommand{\hG}{{\hat G}}

\newcommand{\hu}{{\hat u}}
\newcommand{\hU}{{\hat U}}
\newcommand{\hW}{{\hat W}}
\newcommand{\htV}{{\hat{\tilde V}}}

\newcommand{\De}{\Delta}
\newcommand{\de}{\delta}
\newcommand{\eps}{\epsilon}
\newcommand{\ka}{\kappa}
\newcommand{\la}{\lambda}
\newcommand{\lp}{\lambda_+}
\newcommand{\lm}{\lambda_-}
\newcommand{\La}{\Lambda}
\newcommand{\mum}{\mu_-}
\newcommand{\mup}{\mu_+}

\newcommand{\sg}{\sigma}

\newcommand{\Om}{\Omega}
\newcommand{\om}{\omega}

\newcommand{\omp}{\om_+}

\newcommand{\ze}{\zeta}

\newcommand{\ga}{\gamma}
\newcommand{\gap}{\gamma_+}
\newcommand{\gam}{\gamma_-}
\newcommand{\gappr}{\gamma'_+}
\newcommand{\gampr}{\gamma'_-}

\newcommand{\Ga}{\Gamma}

\newcommand{\dd}{\partial}

\newcommand{\bfo}{{\bf 1}}

\begin{document}

\title[Alternative models for FX and double barrier options in L\'evy models]
{Alternative models for FX, arbitrage opportunities  and efficient pricing of double barrier options in L\'evy models}
\author[
Svetlana Boyarchenko and
Sergei Levendorski\u{i}]
{
Svetlana Boyarchenko and
Sergei Levendorski\u{i}}

\begin{abstract}
We analyze the qualitative differences between prices of double barrier no-touch options
in the Heston model and pure jump KoBoL model calibrated to
the same set of the empirical data, and discuss the potential for arbitrage opportunities if the correct model
is a pure jump model. We explain and demonstrate with numerical examples that accurate and fast calculations of prices of double barrier options in jump models are extremely difficult using the numerical methods available in the literature.
We develop a new efficient method (GWR-SINH method) based of the Gaver-Wynn-Rho acceleration applied to the Bromwich integral;
the SINH-acceleration and simplified trapezoid rule are used to evaluate perpetual double barrier options for each value of the spectral parameter in GWR-algorithm. 
 The program in Matlab running on a Mac with moderate characteristics achieves the precision of the order of E-5 and better in several several dozen of milliseconds; the precision E-07 is achievable in about  0.1 sec.   
We outline the extension of GWR-SINH method to regime-switching models and models with stochastic parameters and stochastic interest rates.

\end{abstract}

\thanks{
\emph{S.B.:} Department of Economics, The
University of Texas at Austin, 2225 Speedway Stop C3100, Austin,
TX 78712--0301, {\tt sboyarch@utexas.edu} \\
\emph{S.L.:}
Calico Science Consulting. Austin, TX.
 Email address: {\tt
levendorskii@gmail.com}}

\maketitle



\tableofcontents

\section{Introduction}\label{s:intro}

There exists a large body of literature devoted
to 
calculation of the expectation of a function of a L\'evy process and its running extremum, and related optimal stopping problems,
standard examples being  barrier and American options, and lookback options with barrier and/or American features. However,
sufficiently accurate and fast algorithms are difficult to develop unless a process is of a rather simple nature. The errors can be especially large for prices of double barrier options in L\'evy and regime-switching L\'evy models.
The importance of a reliable and fast pricing procedure for barrier options in regime-switching L\'evy models is apparent in the following
situation.
 Typically, the best fit to empirically observed  vanilla prices (or implied volatilities) is achieved with jump-diffusion models, and, among models with small number of parameters, with purely jump models. In fact, in many cases,  finite variation L\'evy processes 
 are calibrated better than the processes of infinite variation (see, e.g., examples in \cite{CGMY,ChenFengLin,levendorskii-xie-Asian,AsianGamma}). However,
as  an extensive empirical study in \cite{dahlbokum07} demonstrates, the prices of barrier options provided by Markit agree 
better with prices in diffusion models. If the latter prices are the real prices, the market prices vanillas and barrier options
using non-equivalent probability measures, hence, there may exist arbitrage opportunities. Using the well-known criterion of the equivalency of L\'evy measures (see., e.g., \cite[Thm. 33.1]{sato})
one can easily prove that in several empirical examples documented in \cite{CGMY},  the historic and risk-neutral measures inferred from the prices of a stock and vanillas are not equivalent. The differences among prices of vanillas in many models being small, the apparent violations of the no-arbitrage condition 
can be explained by the transaction cost.

 In the case of barrier options, the differences in prices can be quite substantial due
to boundary effects. The asymptotic analysis  in  \cite{NG-MBS,early-exercise,BIL,asymp-sens} for single barrier options shows that, in many cases, close to the barrier, the prices  in purely jump models are much larger than the prices in models with a sizable diffusion component. The numerical examples in \cite{BIL}  demonstrate that the difference is sizable at the distance of several percent from the barrier. 
Using the fairly complicated but exact analytical formula for prices of the double barrier options in terms of the inverse Laplace-Fourier transform (this formula is implicitly used in the method of the paper), one can derive similar but less explicit asymptotic formulas for double barrier options and conclude that, in the case of the double barrier options,
the differences between prices in purely jump models and models with  sizable diffusion components must be rather large, especially if
the distance between the barriers is fairly small, e.g., 10\%. Even if a high transaction cost is taken into account, the difference
between the prices in the models used by the counterparties to price  double barrier options may create arbitrage opportunities.
The simplest way to realize the arbitrage opportunity in this setting is to buy a double-no-touch options (DNT) from a bank and keep it until maturity. If the discounted probability
of the underlying remaining in the corridor between the two barriers until maturity, calculated in a jump model, is larger than the ask price,
possibly, calculated at the bank using the Heston model, then one has an apparent arbitrage opportunity. 
The first aim of the paper is to analyze this opportunity using a  small set of the real data; the second  aim of the paper
is to analyze analytical difficulties for accurate pricing double barrier options in jump models, and develop an efficient method
for evaluation of double barrier options in one factor L\'evy models. 
The algorithm developed in the present paper is much more accurate and dozens of times faster
than the one in \cite{BLdouble}; the main block of the algorithm is the realization in the dual space of the main block in \cite{BLdouble},
where the calculations are in the state space.
We apply the Gaver-Wynn-Rho algorithm (GWR algorithm) and, for each value of the (positive) spectral parameter,  apply the same algorithm as in
\cite{BLdouble} for perpetual double barrier options but do the calculations in the dual space. The conformal deformation technique
(sinh-acceleration) developed in \cite{SINHregular,Contrarian,EfficientLevyExtremum,EfficientDoubleBarrier} allows one to evaluate the
prices of the perpetual options with the accuracy E-09 and better using arrays of the length of a hundred and shorter. The error tolerance
of the order of E-14 requires arrays several hundreds long. We call the method in the paper the GWR-SINH method.
Since the analytic properties of the characteristic exponent of L\'evy processes considered in the present paper are significantly worse
that the properties of L\'evy processes considered in \cite{EfficientDoubleBarrier}, the Laplace transform of the price of  barrier options
does not admit analytic continuation to the complex plane with the cut of the form $(-\infty, a]$ as in \cite{EfficientDoubleBarrier}.
In the result, the error of the GWR method in this paper is 1-2 orders of magnitude larger than that in  \cite{EfficientDoubleBarrier},
which agrees with   the general characterization of the accuracy of the Gaver method in \cite{AbateValko04,AbValko04b}.


The rest of the paper is organized as follows.
In Sect. \ref{s:auxil}, we collect the definitions of classes of L\'evy processes used in the paper and necessary basic facts of
the Wiener-Hopf factorization. In Sect. \ref{s:double_state}, we recall the scheme of evaluation of the Laplace transform
\cite{BLdouble} using the calculations in the state space. GWR-SINH method and the explicit algorithm are in Sect. \ref{s:GWR-SINH}.
In Sect. \ref{s:pricing_barr_Levy}, we discuss several popular approaches to pricing barrier options and analyze common sources  of errors
of various groups of methods. In Sect. \ref{s:concl}, we summarize the results of the paper and outline the extension of the GWR-SINH method to 
regime-switching models and models with the stochastic volatility and interest rates. Calibration results and prices of DNT options in the Heston and KoBoL models are in Section \ref{s: calib_and_DNT}. The  GWR-SINH method for regime-switching L\'evy models is in Section \ref{s:reg-switch}.


\section{Auxilliary results}\label{s:auxil}
 Let $X$ be a one-dimensional L\'evy process on the filtered probability space $(\Om, \cF, \{\cF_t\}_{t\ge 0}, \bQ)$
satisfying the usual conditions; the riskless rate is constant and $\bQ$ is an equivalent martingale measure.
  We denote the expectation operator under $\bQ$ by $\bE$. The underlying is modeled as $S_t=e^{X_t}$,
  the barriers are $H_-<H_+$, and the maturity date is denoted $T$. We set $h_\pm=\ln H_\pm$ and $x=\ln S$. 
  The infimum process  and supremum
 process $\barX_t=\sup_{0\le s\le t}X_s$ and $\uX_t=\inf_{0\le s\le t}X_s$ are defined pathwise, a.s.
 For $h\in \bR$,  $\tau^+_h$ and $\tau^-_h$ denote the first entrance time of $X$ into $[h,+\infty)$ and
 $(-\infty, h]$, respectively.  For $q>0$,  $T_q\sim\operatorname{Exp}q$ denotes an
exponentially distributed random variable with mean $q^{-1}$, independent of $X$.
 $Q_{GWR}$ denotes the set of nodes  in the GWR algorithm used in Sections \ref{s:double_state} and
\ref{s:GWR-SINH}.

 \subsection{Wiener-Hopf factorization}\label{ss:WHF}
 This subsection contains basic formulas and results systematically used in a number of publications,
 e.g., \cite{NG-MBS,barrier-RLPE,paired,EfficientDiscExtremum,EfficientDoubleBarrier}. 
In probability, the 
Wiener-Hopf factors are defined as
\begin{equation}\label{defphipm}
\phi^+_q(\xi)=\bE[e^{i\xi \barX_{T_q}}],\quad \phi^-_q(\xi)=\bE[e^{i\xi \uX_{T_q}}].
\end{equation}
Functions $\phi^\pm_q(\xi)$ appear in the Wiener-Hopf factorization formula
\begin{equation}\label{whfprob}
\frac{q}{q+\psi(\xi)}=\phi^+_q(\xi)\phi^-_q(\xi),\quad \xi\in\bR.
\end{equation}
Define the expected present value operators (EPV-operators) under $X$, $\barX$ and $\uX$ (all three start at 0) by 
$\cE_q u(x)=\bE\left[u(x+X_{T_q})\right]$, $\cE^+_q u(x)=\bE\left[u(x+\barX_{T_q})\right]$ and
$\cE^-_q u(x)=\bE\left[u(x+\uX_{T_q})\right]$. The EPV operators are bounded operators in $L_\infty(\bR)$.
In the case of L\'evy processes
with exponentially decaying L\'evy densities, the EPV operators are bounded operators in spaces with exponential weights. The operator version of \eqref{whfprob} 
\begin{equation}\label{operWHF}
\cE_q=\cE^+_q\cE^-_q=\cE^-_q\cE^+_q
\end{equation}
is a special case of the operator form of the Wiener-Hopf factorization in the theory of boundary problems for
differential and pseudo-differential operators (pdo). Indeed,  $\cE^\pm_q e^{ix\xi}=\phi^\pm_q(\xi) e^{ix\xi}$.   
 This means that
 $\cE^\pm_q$ are pdo with symbols $\phi^\pm_q$, and $\cE^\pm_qu(x)=\cF^{-1}_{\xi\to x}\phi^\pm_q(\xi)\cF_{x\to\xi}u(x)$ for sufficiently regular functions $u$.  See e.g., \cite{eskin,NG-MBS}.
 The Wiener-Hopf factor $\phi^+_q(\xi)$ (resp., $\phi^-_q(\xi)$) admits analytic continuation to
the half-plane $\{\Im\xi>0\}$ (resp., $\{\Im\xi<0\}$).

The characteristic exponents of all popular classes of L\'evy processes
bar stable L\'evy processes admit analytic continuation to a strip around the real axis. See \cite{NG-MBS,barrier-RLPE,BLSIAM02}, where the general class of Regular L\'evy processes of
exponential type (RLPE) is introduced. 
Let $X$ be a L\'evy process with the characteristic exponent admitting analytic continuation to
a strip $\{\Im\xi\in (\mum,\mup)\}$ around the real axis, and let $q>0$. Then (see., e.g., \cite{NG-MBS,barrier-RLPE,paired})
the following statements hold.
\mbr\noindent
I.  There exist
$\sg_-(q)<0<\sg_+(q)$ such that
\begin{equation}\label{crucial}
q+\psi(\eta)\not\in (-\infty,0],\quad \Im\eta\in (\sg_-(q),\sg_+(q)).
\end{equation}
\mbr\noindent
II. The Wiener-Hopf factor $\phi^+_q(\xi)$ admits analytic continuation
to the half-plane $\{\Im\xi>\sg_-(q)\}$, and can be calculated as follows: for any $\om_-\in (\sg_-(q), \Im\xi)$,
\begin{eqnarray}\label{phip1}
\phi^+_q(\xi)&=&\exp\left[\frac{1}{2\pi i}\int_{\Im\eta=\om_-}\frac{\xi \ln (1+\psi(\eta)/q)}{\eta(\xi-\eta)}d\eta\right].
\end{eqnarray}
\mbr\noindent
III. The Wiener-Hopf factor $\phi^-_q(\xi)$ admits analytic continuation
to the half-plane $\{\Im\xi<\sg_+(q)\}$, and can be calculated as follows: for any $\om_+\in (\Im\xi, \sg_+(q))$,
\begin{eqnarray}\label{phim1}
\phi^-_q(\xi)&=&\exp\left[-\frac{1}{2\pi i}\int_{\Im\eta=\om_+}\frac{\xi \ln (1+\psi(\eta)/q)}{\eta(\xi-\eta)}d\eta\right].
\end{eqnarray}
We can (and will) use \eqref{whfprob} and \eq{phip1} to calculate $\phimq(\xi)$ on $\{\Im\xi\in (0,\sg^+_q)\}$, 
and \eqref{whfprob} and \eq{phim1} to calculate $\phipq(\xi)$ on $\{\Im\xi\in (\sg^-_q,0)\}$.

The explicit algorithm formulated and used in \cite{EfficientDoubleBarrier} for processes of infinite variation
and finite variation processes without drift uses the representations \eq{phip1}-\eq{phim1}. 
In the case of finite variation processes with positive drift, which we consider in the paper, the following set of formulas derived in \cite{NG-MBS,paired} is more efficient:
 \beqa\label{phipq1f}
\phi^{+,0}_q(\xi)&=&\exp\left[-\frac{1}{2\pi i}\int_{\Im\eta=\om_-}\frac{\xi\ln(1+\psi^0(\eta)/(q-i\mu\eta))}{\eta(\eta-\xi)}d\eta\right],\\\label{phipq1ff}
\phipq(\xi)&=&\frac{q}{q-i\mu\xi}\phi^{+,0}_q(\xi),
\\\label{phimq1f}
\phimq(\xi)&=&\exp\left[\frac{1}{2\pi i}\int_{\Im\eta=\om_+}\frac{\xi\ln(1+\psi^0(\eta)/(q-i\mu\eta))}{\eta(\eta-\xi)}d\eta\right].
\eqa
The formulas for the case of finite variation processes with negative drift are by symmetry.

 \subsection{General classes of L\'evy processes amenable to efficient calculations}\label{ss:gen_eff_Levy}
Essentially all popular classes  of L\'evy processes processes enjoy additional properties formalized in 
\cite{SINHregular,EfficientAmenable} as follows. 
For $\nu=0+$ (resp., $\nu=1+$), set $|\xi|^\nu=\ln|\xi|$ (resp., $|\xi|^\nu=|\xi|\ln|\xi|$), and introduce the following complete ordering in
the set $\{0+,1+\}\cup (0,2]$: the usual ordering in $(0,2]$; $\forall\ \nu>0, 0+<\nu$; $\forall\ \nu>1, 1<1+<\nu$.
For $\ga\in (0,\pi]$, $\gap\in (0,\pi/2]$, $\gam\in [-\pi/2,0)$ and $\mum<\mup$, define
 $\cC_{\gam,\gap}=\{e^{i\varphi}\rho\ |\ \rho> 0, \varphi\in (\gam,\gap)\cup (\pi-\gap,\pi-\gam)\}$, 
 $\cC_{\ga}=\{e^{i\varphi}\rho\ |\ \rho> 0, \varphi\in (-\ga,\ga)\}$, $S_{(\mum,\mup)}=\{\xi\ |\ \Im\xi\in (\mum,\mup)\}$.

\begin{defin}\label{def:SINH_reg_proc_1D0}(\cite[Defin 2.1]{EfficientAmenable})
 We say that $X$ is a SINH-regular L\'evy process  (on $\bR$) of   order
 $\nu$ and type $((\mum,\mup);\cC; \cC_+)$
 iff
the following conditions are satisfied:
\begin{enumerate}[(i)]
\item
 $\nu\in\{0+,1+\}\cup (0,2]$ and either $\mum<0\le \mup$ or $\mum\le 0<\mup$;
\item
$\cC=\cC_{\gam,\gap}, \cC_+=\cC_{\gampr,\gappr}$, where $\gam<0<\gap$, $\gam\le \gampr\le 0\le \gappr\le \gap$,
and $|\gampr|+\gappr>0$; 
\item
the characteristic exponent $\psi$ of $X$ can be represented in the form
\bbe\label{eq:reprpsi}
\psi(\xi)=-i\mu\xi+\psi^0(\xi),
\ee
where $\mu\in\bR$, and 
$\psi^0$ admits analytic continuation to $i(\mum,\mup)+ (\cC\cup\{0\})$;
\item
for any $\varphi\in (\gam,\gap)$, there exists $c_\infty(\varphi)\in \bC\setminus (-\infty,0]$ s.t.
\begin{equation}\label{asympsisRLPE}
\psi^0(\rho e^{i\varphi})\sim  c_\infty(\varphi)\rho^\nu, \quad \rho\to+\infty;
\end{equation}
\item
the function $(\gam,\gap)\ni \varphi\mapsto c_\infty(\varphi)\in \bC$ is continuous;
\item
for any $\varphi\in (\gampr, \gappr)$, $\Re c_\infty(\varphi)>0$.
\end{enumerate}
\end{defin}
To simplify the constructions in the paper, we assume that $\mum<0<\mup$ and $\gampr<0<\gappr$. 
\begin{example}\label{ex:KoBoL}{\rm  In \cite{genBS,KoBoL}, we constructed a family of pure jump processes
generalizing the class of  \cite{koponen},  with the L\'evy measure
\begin{equation}\label{KBLmeqdifnu}
F(dx)=c_+e^{\lm x}x^{-\nu_+-1}\bfo_{(0,+\infty)}(x)dx+
 c_-e^{\lp x}|x|^{-\nu_--1}\bfo_{(-\infty,0)}(x)dx,
\end{equation}
where $c_\pm>0, \nu_\pm\in [0,2), \lm<0<\lp$. Starting with \cite{NG-MBS}, we use the name KoBoL processes. If $\nu_\pm\in (0,2), \nu_\pm\neq 1$,
\bbe\label{KBLnupnumneq01}
\psi^0(\xi)=c_+\Ga(-\nu_+)((-\lm)^{\nu_+}-(-\lm-i\xi)^{\nu_+})+c_-\Ga(-\nu_-)(\lp^{\nu_-}-(\lp+i\xi)^{\nu_-}).
\ee
 A specialization
 $\nu_\pm=\nu\neq 1$, $c=c_\pm>0$, of KoBoL used in a series of numerical examples in \cite{genBS} was named CGMY model in \cite{CGMY} (and the labels were changed:
 letters $C,G,M,Y$ replace the parameters $c,\nu,\lm,\lp$ of KoBoL):
 \bbe\label{KBLnuneq01}
 \psi^0(\xi)= c\Ga(-\nu)[(-\lm)^{\nu}-(-\lm- i\xi)^\nu+\lp^\nu-(\lp+ i\xi)^\nu].
\ee
Evidently, $\psi^0$ given by \eq{KBLnuneq01} is analytic in $\bC\setminus i((-\infty,\lm]\cup [\lp,+\infty))$, and  $\forall\ \varphi\in (-\pi/2,\pi/2)$, \eq{asympsisRLPE} holds with
\bbe\label{ascofnupeqnumcc}
c_\infty(\varphi)=-2c\Ga(-\nu)\cos(\nu\pi/2)e^{i\nu\varphi}.
\ee
}
\end{example} 
In \cite{EfficientAmenable}, we defined a class of Stieltjes-L\'evy processes (SL-processes).  Essentially, $X$ is called a (signed) SL-process if $\psi$ is of the form
\bbe\label{eq:sSLrepr}
\psi(\xi)=(a^+_2\xi^2-ia^+_1\xi)ST(\cG^0_+)(-i\xi)+(a^-_2\xi^2+ia^-_1\xi)ST(\cG^0_-)(i\xi)+(\sg^2/2)\xi^2-i\mu\xi, 
\ee
where $ST(\cG)$ is the Stieltjes transform of the (signed) Stieltjes measure $\cG$,  $a^\pm_j\ge 0$, and $\sg^2\ge0$, $\mu\in\bR$.
A (signed) SL-process is called regular if it is SINH-regular. We proved in \cite{EfficientAmenable} that 
the characteristic exponent $\psi$ of 
a (signed) SL-process admits analytic continuation to the complex plane with two cuts along the imaginary axis.
If $X$ is an SL-process, then, for any $q>0$, equation $q+\psi(\xi)=0$ has no solution on $\bC\setminus i\bR$.
We proved that all popular classes of L\'evy processes bar the Merton model and Meixner processes are regular SL-processes, with $\ga_\pm=\pm \pi/2$;
the Merton model and Meixner processes are regular signed SL-processes, and $\ga_\pm=\pm \pi/4$.
In \cite{EfficientAmenable}, the reader can find a list of SINH-processes and SL-processes, with calculations of the order and type.


\section{Double barrier options: calculations in the state space}\label{s:double_state}
 \subsection{Laplace transform and its inversion}
 Let $r\in \bR$, $T>0$, $h_-<x<h_+$, and  $G\in L_\infty((h_-,h_+))$. 
  To evaluate
 \bbe\label{Vdoublent1}
 V(G;r; h_-,h_+;T,x)=\bE^x[e^{-rT}\bfo_{\tau^-_{h_+}\wedge \tau^+_{h_+}>T}G(X_T)],
 \ee 
 we take $r_0\in \bR$ and represent $V(G; r; h_-,h_+;T,x)$ in the form $V(G;h_-,h_+;T,x)=e^{rT}
 V(G; r+r_0; h_-,h_+;T,x)$. First, we apply the Laplace transform w.r.t. $T$ to $V(G; r+r_0; h_-,h_+;T,x)$. 
 Then, assuming that $r+r_0>0$ is sufficiently large so that $
 \tV(G; r+r_0; h_-,h_+;q',x)$ can be efficiently calculated for $q'$ in the right half-palne,
 we  evaluate the Bromwich integral 
 \bbe\label{inv_tV0}
  V(G; r+r_0; h_-,h_+;T,x)=\frac{1}{2\pi i}\int_{\Re q'=\sg}e^{q'T} \tV(G; r+r_0; h_-,h_+;q',x),
  \ee
  where $\sg>0$ is arbitrary,
using an appropriate numerical method. Finally, we multiply the result by $e^{r_0T}$. 
In the case $\nu\in (0,1)$ and $\mu\neq 0$,
the most efficient method of the Laplace inversion of complicated functions, namely, the sinh-acceleration
used in \cite{EfficientDoubleBarrier}, is not applicable. Therefore, we use the GWR algorithm.
Let $Q_{GS}$ be the set of spectral parameters used in the GWR algorithm. We need to calculate the prices of perpetual options $\tV(G; h_-,h_+;q,x)$ for $q\in r+r_0+Q_{GS}$. The first advantage of using $r_0$ is that we can ensure that $q+\psi(\xi)\not\in (-\infty,0]$  for
all $q,\xi$ of interest, which simplifies the verification of several important technical conditions in the paper, starting with conditions for an efficient
evaluation of the Wiener-Hopf factors. The second advantage is  that the condition $q+\psi(\xi)\not\in (-\infty,0]$ allows for deformations of the contours of integration in
the formulas for the Wiener-Hopf factors and $\tV(G; h_-,h_+;q'+r+r_0,x)$ that make it possible to evaluate the prices  of the perpetual options with the precision E-10 and better. The third advantage stemming from the second one is that if the differences between prices of the option with the finite time horizon calculated for different $r_0$'s 
are of the order of E-05, then E-05 is the order of the error of the GWR method itself. The disadvantage is that if 
$T$ is large,
and the calculations are with double precision, then  the factors $e^{-r_0T}$ and $e^{r_0T}$ may
 introduce large errors. This difficulty can be resolved using an additional trick in \cite{paired}.  However, in the case of double barrier options
 with small distances between barriers, the price of options of long maturity is negligible, hence, we may assume that $T$ is not large.

  \subsection{Evaluation of the perpetual double barrier options}
The notation and scheme in this Section are borrowed from \cite{BLdouble,EfficientDoubleBarrier};
 we need the scheme as the starting point for the scheme where the calculations are in the dual space.
 Let $q\in r+r_0+Q_{GS}$ and
  $x\in (h_-,h_+)$. We have
    \bbe\label{eq:repr1}
\tV(G; h_-,h_+;q,x)=q^{-1}\bE^x[\bfo_{\tau^-_{h_+}\wedge \tau^+_{h_+}>T_q}G(X_{T_q})].
 \ee
 Let $lG$ be a bounded measurable extension of $G$ to $\bR$. 
Set $\tV^0(lG; q,\cdot)=q^{-1}\cEq lG$, and calculate $\tV^1(lG;h_-,h_+;q,x):=\tV(G;h_-,h_+;q,x)-\tV^0(lG; q,x)$
 as the sum of a series exponentially converging in $L_\infty$-norm. 
The terms of the series depend on the choice of the extension but $\tV(G;h_-,h_+;q,x)$ is independent of
the choice.  
Define
 \beqa\label{def:tVp1}
\tV^+_1(lG;h_-,h_+;q,x)=\bE^x[e^{-q\tau^+_{h_+}}\tV^0(lG;q, X_{\tau^+_{h_+}})],\
\\
\label{def:tVm1}
\tV^-_1(lG;h_-,h_+;q,x)=\bE^x[e^{-q\tau^-_{h_-}}\tV^0(lG;q, X_{\tau^-_{h_-}})],
\eqa
and note that  $\tV^+_1$ (resp., $\tV^-_1$) is the  EPV of the stream $G(X_t)$ which starts to accrue the first time $X_t$ crosses $h_+$ from below
(resp., $h_-$ from above). Inductively, for $j=2,3,\ldots,$ define
\beqa\label{def:tVpj}
\tV^+_j(lG;h_-,h_+;q,x)=\bE^x[e^{-q\tau^+_{h_+}}\tV^-_{j-1}(lG;h_-,h_+;q, X_{\tau^+_{h_+}})],\
\\
\label{def:tVmj}
\tV^-_j(lG;h_-,h_+;q;x)=\bE^x[e^{-q\tau^-_{h_-}}\tV^+_{j-1}(lG;h_-,h_+;q, X_{\tau^-_{h_-}})].
\eqa
In \cite{BLdouble}, the following key theorem is derived from the stochastic continuity of $X$.
\begin{thm}\label{thm:convergence}
For any $\sg>0$ and $h_-<h_+$, there exist $\de_\pm=\de_\pm(\sg, h_+-h_-)\in (0,1)$ such that for all 
$q\ge \sg$, $lG\in L_\infty(\bR)$ and $j=2,3,\ldots$, 
\beqa\label{eq:boundtVp}
\sup_{x\ge h_+}|\tV^+_j(lG;h_-,h_+;q;x)|&\le&\de_+\sup_{x\le h_-}|\tV^-_{j-1}(lG;h_-,h_+;q,x)|,\\
\label{eq:boundtVm}
\sup_{x\le h_-}|\tV^-_j(lG;h_-,h_+;q;x)|&\le&\de_- \sup_{x\ge h_+}|\tV^+_{j-1}(lG;h_-,h_+;q,x)|,
\eqa
and
\beqa\label{qtV}
 \tV^1(lG;h_-,h_+;q,x)&=&\sum_{j=1}^{+\infty} (-1)^j (\tV^+_j(lG;h_-,h_+;q,x)+\tV^-_j(lG;h_-,h_+;q,x)).
 \eqa
 The series on the RHS of \eq{qtV} exponentially converges in $L_\infty$-norm.

\end{thm}

Under additional weak conditions on $X$, Theorems 11.4.4 and 11.4.5 in \cite{IDUU} state that 
\beqa\label{tVhm1}
\tV^-_1(lG; h_-,h_+; q,x)&=&q^{-1}(\cEmq\bfo_{(-\infty,h_-]}\cEpq)lG)(x),\\
\label{tVhp1}
\tV^+_1(lG; h_-,h_+; q,x)&=&q^{-1}(\cEpq\bfo_{[h_+,+\infty)}\cEmq)lG)(x);
\eqa
in \cite{single}, \eq{tVhm1}-\eq{tVhp1} are proved for any L\'evy process.
Similar representations for $\tV^\mp_j$, $j=2,3,\ldots$:
\beqa\label{tVhpj2}
\tV^+_j(lG; h_-,h_+; q,x)&=&(\cE^+_q\bfo_{[h_+,+\infty)}(\cE^+_q)^{-1})\tV^-_{j-1}(lG;h_-,h_+; q,x),
\\\label{tVhmj2}
\tV^-_j(lG;h_-,h_+; q,x)&=&(\cE^-_q\bfo_{(-\infty, h_-]}(\cE^-_q)^{-1}\tV^+_{j-1})(lG;h_-,h_+; q,x),
\eqa
follow from Theorems 11.4.6 and 11.4.7 in \cite{IDUU}. See \cite{EfficientDoubleBarrier} for details.

\begin{rem}\label{rem:convergence}{\rm 
\begin{enumerate}[(a)]
\item
  For a numerical realization, the series on the RHS of \eq{qtV} are truncated,
and $\sum_{j=1}^\infty$ replaced with $\sum_{j=1}^{M_0}$; given the error tolerance,
 $M_0$ can be chosen using the bounds in Theorem \ref{thm:convergence}.

\item If $G$ is bounded on $[h_-,h_+]$ but given by an analytical expression that defines an unbounded function on $\bR$, we can use the scheme above replacing $G$ with a function $lG$ of class $L_\infty(\bR)$, which coincides with $G$ on $[h_-,h_+]$. This simple consideration suffices for a numerical realization  in the state space.
If the numerical realization is in the dual space, complex-analytical properties of the Fourier transform $\hG$ are crucial.
If $\hG$ has good properties as in the case of the call option, then a good replacement  of $G$   with a bounded function
requires additional work. Instead, we will not replace $G$ but assume that the strip of analyticity of $\psi$ is sufficiently wide, and
$q>0$ is sufficiently large so that the series \eq{qtV} converges in a space with an appropriate exponential weight. 

\item Alternatively, in the case of a call option, one can make an appropriate Esscher transform and reduce to the case of 
a bounded $G$.

\item
The inverse Laplace transform of $\tV^0(q,x)=q^{-1}\cEq G(x)$ is the price $V_{\mathrm{euro}}(G; T,x)$ of the European option
 with the payoff $G(x+X_T)$ at maturity date $T$. An explicit procedure for an efficient numerical evaluation of $V_{\mathrm{euro}}(G; T,x)$
 can be found in \cite{SINHregular}. In the paper, we design an efficient numerical procedure for the evaluation of 
 $V^1(G; h_-,h_+;T,x)= V(G; h_-,h_+;T,x)-V_{\mathrm{euro}}(G; T,x)$. Note that  $V^1(G; h_-,h_+;T,x)$ is the inverse Laplace transform of the series
 on the RHS of \eq{qtV}.
 \item
 To shorten the notation, we suppress the dependence of $\tV^0, \tV^1$ and $\tV^\pm_j$ on $lG$.
 \end{enumerate}
}
\end{rem}


\section{GWR-SINH method}\label{s:GWR-SINH}
\subsection{Calculation of the Wiener-Hopf factors for $q>0$. The case $\mu>0$ and $\nu\in (0,1)$}
Set $q_0=\min Q_{GWR}$ .  First, we construct the curves $\cL^+_{\om_{1,+}, b_+,\om_+}$ and
$\cL^-_{\om_{1,-}, b_-,\om_-}$ in the upper and lower half-planes, respectively, such that 
\begin{enumerate}[(1)]
\item
the curve $\cL^+_{\om_{1,+}, b_+,\om_+}$ intersects the imaginary axis at a point $ia_+$,
where $a_+\in (0, \lp)$, and $q_0+\psi(ia_+)>0$; 
\item
the curve $\cL^-_{\om_{1,-}, b_-,\om_-}$ intersects the imaginary axis at a point $ia_-$,
where $a_-\in (\lm,0)$, and $q_0+\psi(ia_-)>0$ and $q_0+a_-\mu>0;$
\item
for all $\eta\in \cL^+_{\om_{1,+},b_+,\om_+}\cup\cL^-_{\om_{1,-},b_-,\om_-}$, $q_0+\psi(\eta)\not\in (-\infty,0]$. 
\end{enumerate} 
We deform the contours of integration in \eq{phipq1f} and \eq{phimq1f} and  calculate \beqast\label{phipq1fsinh}
\phi^{+,0}_q(\xi)&=&\exp\left[-\frac{1}{2\pi i}\int_{\cL^-_{\om_{1,-}, b_-,\om_-}}\frac{\xi\ln(1+\psi^0(\eta)/(q-i\mu\eta))},
\ \xi\in {\eta(\eta-\xi)}d\eta\right],\ \cL^+_{\om_{1,+}, b_+,\om_+},
\\\label{phipq1fsinh2}
\phipq(\xi)&=&(1-i\mu\xi/q)^{-1}\phi^{+,0}_q(\xi), \ \cL^+_{\om_{1,+}, b_+,\om_+},
\\\label{phimq1fsinh}
\phimq(\xi)&=&\exp\left[\frac{1}{2\pi i}\int_{\cL^+_{\om_{1,+}, b_+,\om_+}}\frac{\xi\ln(1+\psi^0(\eta)/(q-i\mu\eta))}{\eta(\eta-\xi)}d\eta\right], \ \xi\in\cL^-_{\om_{1,-}, b_-,\om_-}.
\eqast
To calculate $\phipq(\xi)$, $\xi\in\cL^-_{\om_{1,-}, b_-,\om_-}$, and $\phimq(\xi)$, $\xi\in\cL^+_{\om_{1,+}, b_+,\om_+}$, we use \eq{whfprob}.
\begin{rem}\label{rem:choice}{\rm
In typical cases, $\mu$ and $c\Ga(-\nu)$ are of the order of $1$ or even 0.1, and $\lp$ and $-\lm$ are greater than 3.
Therefore, it is possible to satisfy (1) with $a_+>2$ and (2) using $a_-<-1$. Typically, the choice of large $a_\pm$ (in absolute value)
  is not optimal, and,
 in this paper, we choose $a_\pm=\pm 1$,  $\om_{1,\pm}=0$, $\om_\pm=\pm \pi/4$, $d_\pm=\pi/4$,
and set $b_\pm=1/\sin(\omp+d_+)=1$ (or smaller multiplying by, e.g.  $0.9$). In the program, we first check that, with this choice, the conditions (1) and (2) are satisfied, and then check condition (3).
}
\end{rem}

\subsection{General scheme for a fixed $q>0$}\label{ss:gen_q>0} 
The numerical scheme in this section has  small but crucial differences from
 the scheme in \cite{EfficientDoubleBarrier}. Writing the program, the reader can check that the scheme  formulated in 
 \cite{EfficientDoubleBarrier} blows up for processes of finite variation with positive drift.  We indicate the changes when they appear.
 For the same of brevity, we formulate the scheme for the case of DNT options. (The reader can easily adjust the schemes for
 call and put options and digitals using the first step of the scheme in \cite{EfficientDoubleBarrier}.)
 With the exception of the last step, when the inverse Fourier transform is applied, the calculations are in the dual space.
 The Fourier transforms $\htV^\pm_j(h_-,h_+;q,\xi)$ of functions $\tV^\pm_j(h_-,h_+;q,x)$ are evaluated 
 on the curves $\cL^\mp:=\cL^\mp_{\om_\mp, b_\mp,\om_{1,\mp}}$ used to evaluate the Wiener-Hopf factors; changing the curves, we can double-check the accuracy of calculations.  In the case of DNT, $\tV^0(q,x)=1/q$, hence, 
\bbe\label{htV1p}
\htV^\pm_1(h_-,h_+;q,\xi)=\pm e^{-ih_\pm\xi}\frac{\phi^\pm_q(\xi)}{i\xi q}, \ \xi\in \cL^\mp.
\ee
For $j=1,2,\ldots, $ define
   \bbe\label{def:hWj}
  \hW_j^\pm(h_-,h_+; q,\xi)=qe^{ih_\pm\xi}\phi^\pm_q(\xi)^{-1}\htV_j^\pm(h_-,h_+;q,\xi), \ \xi\in\cL^\mp.
  \ee
Evidently,
  \bbe\label{tVtW}
  \htV_j^\pm(h_-,h_+;q,\xi)=q^{-1}e^{-ih_\pm\xi}\phi^\pm_q(\xi)\hW^\pm_j(h_-,h_+; q,\xi), \ \xi\in\cL^\mp,
  \ee
and $\hW^\pm_1(h_-,h_+,q,\xi)=\pm 1/(i\xi)=\mp i/\xi$, $\xi\in \cL^\mp$.  For $j=2,3,\ldots,$ $\hW^\pm_j$ are calculated inductively. We rewrite \eq{tVhpj2}-\eq{tVhmj2} in terms of $ \hW^\pm_j$ as follows.
 For $\xi\in \cL^-$, we have
  \beqast
  \hW^+_j(h_-,h_+; q,\xi)&=&qe^{ih_+\xi}\cF_{x\to \xi}\bfo_{[h_+,+\infty)}(\cE^+_q)^{-1}\tV^-_{j-1}(h_-,h_+;q,x)
  \\
  &=& e^{ih_+\xi}\int_{h_+}^{+\infty}dy\, e^{-iy\xi}
  \frac{1}{2\pi}\int_{\cL^+}e^{i(y-h_-)\eta}\frac{\phi^-_q(\eta)}{\phi^+_q(\eta)}\hW^-_{j-1}(h_-,h_+;q,\eta)d\eta\\
  &=&-\frac{e^{ih_+\xi}}{2\pi i}\int_{\cL^+}\frac{e^{-ih_+(\xi-\eta)-ih_-\eta}}{\eta-\xi}
  \frac{\phi^-_q(\eta)}{\phi^+_q(\eta)}\hW^-_{j-1}(h_-,h_+;q,\eta)d\eta
   \eqast
  (we can apply Fubini's theorem because there exists $c>0$ such that $\Im(\eta-\xi)>c|\eta|$ for $\eta\in \cL^+$
  and $\xi\in \cL^-$). Simplifying,
  \bbe\label{hWp}
  \hW^+_j(h_-,h_+; q,\xi)=-\frac{1}{2\pi i}\int_{\cL^+}\frac{e^{i(h_+-h_-)\eta}}{\eta-\xi}
  \frac{\phi^-_q(\eta)}{\phi^+_q(\eta)}\hW^-_{j-1}(h_-,h_+;q,\eta)d\eta,\ \xi\in\cL^-.
  \ee
  Similarly, for $\xi\in \cL^+$, we calculate
  \beqast
  \hW^-_{j}(h_-,h_+; q,\xi)&=& e^{ih_-\xi}\int_{-\infty}^{h_-}dy\, e^{-iy\xi}
  \frac{1}{2\pi}\int_{\cL^-}e^{i(y-h_+)\eta}\frac{\phi^+_q(\eta)}{\phi^-_q(\eta)}\hW^+_{j-1}(h_-,h_+;q,\eta)d\eta\\
  &=&\frac{e^{ih_-\xi}}{2\pi i}\int_{\cL^-}\frac{e^{-ih_-(\xi-\eta)-ih_+\eta}}{\eta-\xi}
  \frac{\phi^+_q(\eta)}{\phi^-_q(\eta)}\hW^+_{j-1}(h_-,h_+;q,\eta)d\eta. 
  \eqast
  Simplifying,
  \bbe\label{hWm}
  \hW^-_{j}(h_-,h_+; q,\xi)=\frac{1}{2\pi i}\int_{\cL^-}\frac{e^{-i(h_+-h_-)\eta}}{\eta-\xi}
  \frac{\phi^+_q(\eta)}{\phi^-_q(\eta)}\hW^+_{j-1}(h_-,h_+;q,\eta)d\eta,\ \xi\in \cL^+.
  \ee
  In the cycle $j=2,3\ldots,$ we calculate  $\hW^\pm_{j}(h_-,h_+; q,\xi)$ and partials sums of the series
   \bbe\label{Winf}
   \hW^\pm(h_-,h_+; q,\xi)=\sum_{j=1}^\infty (-1)^j \hW^\pm_j(h_-,h_+; q,\xi), \ \xi\in \cL^\mp.
   \ee
  Finally,    for $x\in (h_-, h_+)$, we calculate
      \bbe\label{tVqpm}
      \tV^1_\pm(lG;h_-,h_+;T,x)=\frac{1}{2\pi}\int_{\cL^\mp}e^{i(x-h_\pm)\xi}\phi^\pm_q(\xi)\hW^\pm(h_-,h_+; q,\xi)d\xi,
      \ee
      then
   \bbe\label{tVqntfin}
   V^1(lG;h_-,h_+;T,x)=\frac{1}{2\pi i}\int_{\Re q=\sg}dq\,\frac{e^{qT}}{q}(\tV^1_+(lG;h_-,h_+;T,x)+
   \tV^1_-(lG;h_-,h_+;T,x)),
   \ee
   and
   \bbe\label{eq:double_final}
  V(G;h_-,h_+;T,x)=V_{\mathrm{euro}}(lG;T,x)+V^1(lG;h_-,h_+;T,x).
 \ee
 \begin{rem}\label{rem:convergence2}{\rm Theorem \ref{thm:convergence} implies that if 
 the Gaver-Stehfest method or GWR algorithm is used to numerically evaluate
 the Bromwich integral, then the inverse Fourier transform of the series on the RHS of \eq{tVqntfin} converges for any $h_-<x<h_+$
 and any $q>0$ used in the algorithm. It can be proved that the infinite sums of the Fourier transforms converge as well.
 }
\end{rem}
\subsection{
Efficient numerical evaluation of the series \eq{Winf}
}\label{sss:effhWseriesI}
Define operators\\ $\cK_{-+}(=\cK_{-+}(q; \cL^+; h_-,h_+))$ and $\cK_{+-}(=\cK_{+-}(q; \cL^-; h_-,h_+))$ by
\beqa\label{defKmp}
\cK_{-+}\hu(\xi)&=&\frac{1}{2\pi }\int_{\cL^+}\frac{e^{i(h_+-h_-)\eta}(1-i\mu\eta/q)}{\eta-\xi}
  \frac{\phi^-_q(\eta)}{\phi^{+,0}_q(\eta)}\hu(\eta)d\eta,\ \xi\in \cL^-,\\
  \label{defKpm}
\cK_{+-}\hu(\xi)&=&\frac{1}{2\pi }\int_{\cL^-}\frac{e^{-i(h_+-h_-)\eta}}{\eta-\xi}
  \frac{\phi^+_q(\eta)}{\phi^-_q(\eta)}\hu(\eta)d\eta,\ \xi\in \cL^+.
  \eqa
\begin{rem}\label{rem:Kmp_nu<1_mu>0} {\rm In \cite{EfficientDoubleBarrier}, the formula for $\cK_{-+}$ is of the same form as
the one for $\cK_{+-}$: just change the signs. However, in the case $\nu<1$ and $\mu>0$, the ratio $\phi^-_q(\eta)/\phi^+_q(\eta)$
increases as $\eta\to \infty$ very fast, and the algorithm (as in \cite{EfficientDoubleBarrier}) which explicitly uses this ratio
becomes unstable. The factor $e^{i(h_+-h_-)\eta}(1-i\mu\eta/q)$ is uniformly bounded and decays fast at infinity,
and the factor $\phi^-_q(\eta)/\phi^{+,0}_q(\eta)$ is bounded.}
\end{rem}
We write \eq{hWp} and \eq{hWm} as 
  \bbe\label{hWpmj}
  \hW^+_{j+1}=i\cK_{-+}\hW^-_j,\  \hW^-_{j+1}=-i\cK_{+-}\hW^+_j, j=1,2,\ldots
  \ee
  and calculate $\hW^\pm_{j+1}$ and the partial sums in the cycle in $j=1,2,\ldots, M_0$. For the choice
  of the truncation parameter $M_0$
  given the error tolerance,  see Remark \ref{rem:convergence}. This choice is made assuming that
  the total error of calculation of the individual terms is sufficiently small and can be disregarded.
  
  We calculate  $\hW^\pm_{j+1}$ at points of sinh-deformed uniform grids 
  $\xi^\mp_k=i\om^{1,\mp}+b^\mp \sinh(i\om^\mp+y^\mp_k)$, $y^\mp_k=\ze^\mp k$, $k\in \bZ$, 
  on $\cL^\mp$, truncate the grids, and 
  approximate the operators $\cK_{-+}, \cK_{+-}$  with the corresponding matrix operators. The parameters of the conformal deformations
  and corresponding changes of variables and steps $\ze^\pm$ are chosen
  as in \cite{Contrarian,EfficientLevyExtremum,EfficientDoubleBarrier}.   The truncation parameters $N^\pm$ are chosen taking into account the exponential rate of decay of the kernels of the integral operators
  $\cK_{-+}$ and $\cK_{+-}$ w.r.t. the second argument, and exponential decay of $e^{i(x-h_-)\xi}$ as $\xi\to\infty$ along $\cL^+$,
  and $e^{i(x-h_+)\xi}$ as $\xi\to\infty$ along $\cL^-$. In the $y^\pm$-coordinates, the rate of decay is double-exponential,
  hence, the truncation parameters $\La^\pm=N^\pm\ze^\pm$  sufficient to satisfy a small error tolerance $\eps$
  are moderately large. As a simple rule of thumb, we suggest to choose $\La^\pm$ so that
 $ \exp[b^-(x-h_+)\ka_-\sin|\om^-|e^{\La^-}]<\eps,\ \exp[b^+(h_--x)\ka_+\sin(\om^+)e^{\La_+}]<\eps,$
  where $\ka_\pm\in (0,0.5)$, e.g., $\ka_\pm=0.4$. Note that a choice of a smaller $\ka_\pm$, e.g., $\ka_\pm=0.3$,
 does not  increase  $N_\pm$ significantly but makes the prescription more reliable. 
    Keeping the notation $\cK_{-+}$ and  $\cK_{+-}$ for the matrices, we calculate the matrix elements as follows:
  \beqast\label{Kmpjik}
  \cK_{-+}&=&\frac{\ze^+b^+}{2\pi}\left[\frac{e^{i(h_+-h_-)\xi^+_k}(1-i\mu \xi^+_k/q)}{\xi^+_k-\xi^-_j}
  \cdot\frac{\phimq(\xi^+_k)}{\phi^{+,0}_q(\xi^+_k)}\cosh(i\om^++y^+_k)\right]_{|j|\le N^-, |k|\le N^+},
  \\\label{Kpmjik}
  \cK_{+-}&=&\frac{\ze^-b^-}{2\pi}\left[\frac{e^{-i(h_+-h_-)\xi^-_k}}{\xi^-_k-\xi^+_j}
  \cdot\frac{\phipq(\xi^-_k)}{\phimq(\xi^-_k)}\cosh(i\om^-+y^-_k)\right]_{|j|\le N^+, |k|\le N^-}.
  \eqast
\begin{rem}\label{rem:matrix_inv} {\rm If the sizes of matrices $\cK_{-+}$ and  $\cK_{+-}$, hence, 
  the sizes of matrices $\cK^+:=\cK_{-+}\cK_{+-}, \cK^-:=\cK_{+-}\cK_{-+}$ are moderate so that the inverse matrices $(I-\cK^\pm)^{-1}$
  can be efficiently calculated, one can calculate infinite sums using standard matrix tools. See \cite{EfficientDoubleBarrier}. 
  }
  \end{rem} 
  \subsection{Algorithm of GWR-SINH method for DNT options}\label{ss:algo_GWR-SINH_DNT} 
 Steps I-V are preliminary ones;
  Steps VI-IX are performed for each $q\in r+r_0+Q_{GWR}$ used in the GWR method; the calculations can be easily parallelized.
  
   \begin{enumerate}[Step I.]
\item 
Choose $r_0$ and the order of the GWR algorithm, and calculate the set of nodes $Q_{GWR}$ and weights of the GWR algorithm.
\item
{\em Grids for the approximations of $\hW^\pm_m$ and operators $\cK_{-+}, \cK_{+-}, \cK^\pm$}.  Choose  the sinh-deformations and grids for the simplified trapezoid rule on $\cL^\pm$: $\vec{y^\pm}:=\ze^\pm*(-N^\pm:1:N^\pm)$,
$\vec{\xi^\pm}:=i*\om_1^\pm+ b^\pm*\sinh(i*\om^\pm+i\vec{y^\pm})$.  Calculate $\psi^\pm:=\psi(\vec{\xi^\pm})$ and 
$\vec{der^\pm}:=b^\pm*\cosh(i*\om^\pm+\vec{y^\pm}).
$
\item
{\em Grids for evaluation of the Wiener-Hopf factors $\phi^\pm_q(\xi)$.} Choose longer and finer grids for the simplified trapezoid rule on $\cL^\pm_1$: $\vec{y^\pm_1}=\ze_1^\pm*(-N^\pm_1:1:N^\pm_1)$,
$\vec{\xi^\pm_1}:=i*\om^\pm_1+ b^\pm_1*\sinh(i*\om_1^\pm+i\vec{y^\pm_1})$.  Calculate $\psi^{0,\pm}_1:=\psi^0(\vec{\xi^\pm_1})$ and 
$\vec{der^\pm_1}:=b^\pm_1*\cosh(i*\om_1^\pm+\vec{y^\pm_1}).
$ 
\item
Calculate 2D arrays 
\beqast
D^{+-}_1&:=&1./(\mathrm{conj}(\vec{\xi^-_1})'*\mathrm{ones}(1,2*N^++1)-\mathrm{ones}(2*N^-_1+1,1)*\vec{\xi^+})),\\
D^{-+}_1&:=&1./(\mathrm{conj}(\vec{\xi^+_1})'*\mathrm{ones}(1,2*N^-+1)-\mathrm{ones}(2*N^+_1+1,1)*\vec{\xi^-})),\\
D^{+-}&:=&1./(\mathrm{conj}(\vec{\xi^-})'*\mathrm{ones}(1,2*N^++1)-\mathrm{ones}(2*N^-+1,1)*\vec{\xi^+})),\\
D^{-+}&:=&1./(\mathrm{conj}(\vec{\xi^+})'*\mathrm{ones}(1,2*N^-+1)-\mathrm{ones}(2*N^++1,1)*\vec{\xi^-})).
\eqast
\item
{\em Calculate  $\hW^+_1=-i./\vec{\xi^-}, \hW^-_1=i./\vec{\xi^+}$.}
\item
{\em Calculate} $\vec{\phi^{+,0}_q}=\phi^{0,+}_q(\vec{\xi^+})$ and $\vec{\phimq}=\phimq(\vec{\xi^-})$:
\beqast
\Phi^\pm &=&1+\psi^{0,\pm}_1./(q-i*\mu*\vec{\xi^\pm_1}), \\
\vec{\phi^{0,+}_q}&:=&\exp((\ze^-_1*i/(2*\pi))*\vec{\xi^+_1}./.*((\log(\Phi^-).*\vec{\xi^-_1}.*\vec{der^-_1})*D^{-+}_1)),\\
\vec{\phimq}&:=&\exp(-(\ze^+_1*i/(2*\pi))*\vec{\xi^-_1}.*((\log(\Phi^+)*\vec{\xi^-_1}.*\vec{der^+_1}.*D^{+-}_1)),
\eqast
and then
$\phimq(\vec{\xi^+}):=1./\Phi^+./\vec{\phipq},\ \phi^{+,0}_q(\vec{\xi^-}):=1./\Phi^-./\vec{\phimq},$\\ $ \phi^{+}_q(\vec{\xi^-})=\phi^{+,0}_q(\vec{\xi^-})./(1+\psi^{0,-}./(1-i*(\mu/q)*\vec{\xi^-})).
$

\item
{\em Calculate matrices $\cK_{+-}, \cK_{+-}$:}
\beqast
\cK_{-+}&:=&(\ze^+/(2*\pi))*\mathrm{diag}(\vec{der^+}.*(\phimq(\vec{\xi^+})./\phi^{+,0}_q(\vec{\xi^+}))...\\
&& .*(\exp(i*(h_+-h_-)*\vec{\xi^+}).*(1-i*(\mu/q)*\vec{\xi^+}))*D^{-+};\\
\cK_{+-}&:=&(\ze^-/(2*\pi))*\mathrm{diag}(\vec{der^-}.*(\phipq(\vec{\xi^-})./\phimq(\vec{\xi^-})).*\exp(-i*(h_+-h_-)*\vec{\xi^-}))*D^{+-}.\\
\eqast
\item
Use one of the following three blocks. Calculations using Block (1) and either Block (2) or Block (3) can be used to check the accuracy of the result.
\begin{enumerate}[(1)]
\item\begin{itemize}
\item
Assign $\hU^\pm_1=-\hW^\pm_1,$ $\hW^\pm:=\hU^\pm_1$.
\item
In the cycle $j=1,2,\ldots,M_0$, calculate
$
 \hU^+_{2}:=-i*\hU^-_1*\cK_{-+},\\ \hU^-_{2}:=i*\hU^+_1*\cK_{+-},\
 \hU^\pm:=\hW^\pm+\hU^\pm_2, \ \hU^\pm_1:=\hU^\pm_2$.
 \item
 Assign $\hW^\pm=\hU^\pm$.
 \end{itemize}
\item
Calculate 
\begin{itemize}
\item
$\cK^-:=\cK_{-+}*\cK_{+-}, 
\cK^+:=\cK_{+-}*\cK_{-+}$;
\item
 $\hW^+_2:=i*\hW^-_1*\cK_{-+}, \hW^-_2:=-i*\hW^+_1*\cK_{+-}$;
 \item
 $
\hW^{\pm,0}:=\hW^\pm_2-\hW^\pm_1$; 
 \item
 inverse matrices 
$(I-\cK^\pm)^{-1}$;
\item
$\hW^\pm=\hW^{\pm,0}*(I-\cK^\mp)^{-1}$.
\end{itemize}
\item
Replace the last two steps of Block (2) with 
\[
\hW^\pm=\mathrm{conj}(\mathrm{linsolv}(\mathrm{diag}(\mathrm{ones}(2*N^\mp+1))-\mathrm{conj}(\cK^\mp)', \mathrm{conj}(\hW^{\pm,0})'))'.
\]
Typically, the program with  Block (1) achieves precision of the order of E-15 with $M_0=9$ or even $M_0=8$.
The CPU time is several times smaller than with Block (2). Program with Block (2) is faster if
the digitals or vanillas for many strikes need to be calculated. Block (3) is twice faster than Block (2) 
if applied only once.
\end{enumerate}
\item
{\em For $x\in (h_-,h_+)$, calculate} 
\beqast
 V^+&=&(\ze^-/(2*\pi))*\sum(\hW^+.*\exp(i*(x-h_+)*\vec{\xi^-}).*\phipq(\vec{\xi^-}).*\vec{der^-}), \\
V^-&=&(\ze^+/(2*\pi))*\sum(\hW^-.*\exp(i*(x-h_-)*\vec{\xi^+}).*\phimq(\vec{\xi^+}).*\vec{der^+});
\eqast
 this step can be easily parallelized for a given array $\{x_j\}$.
 \item
 Use the GWR algorithm to evaluate $V^1$. \item
{\em Final step.} Set $V=e^{r_0T}(1+V^1)$.
\end{enumerate}

\subsection{A numerical example: Table~\ref{Table 3}}\label{ss:numer_example}  
The calculations   
were performed in MATLAB 2017b-academic use, on a MacPro Chip Apple M1 Max Pro chip
with 10-core CPU, 24-core GPU, 16-core Neural Engine 32GB unified memory,
1TB SSD storage.
The  CPU times shown can be significantly improved using parallelized calculations of the Wiener-Hopf factors and the main block, for each $q$ used
in the Laplace inversion procedure. The parallelization w.r.t.
$H_\pm, S$ is also possible.

\begin{table}
 \caption{\small  Prices of the double no-touch options, and errors (rounded) of Carr's randomization algorithm \cite{BLdouble}
 and GWR-SINH method. Dependence on the spot. Barriers: $H_-=0.95, H_+=1.95$, time to maturity $T=0.25$.
KoBoL model with the parameters in Table \ref{Table 2}, MB: $\nu=0.445, c=1.125,	\lp=	27.93, \lm=	-51.66, \mu=	0.0940$.}
 \begin{tabular}{c|ccccc|c}
 \hline
 $S$ & 0.96	&	0.98	  &	1 &	 1.02 &		1.04 & Time \\
 BB & 0.4325056 &	0.6497429  &	0.6801758  &	0.5289720 &	0.224546 & 812
 \\\hline
$ \eps_0$ & 
 -1.07E-08	 &	1.82E-10	&	-9.46E-09	&	3.68E-09	&	-3.10E-09 & 767.2\\
$ \eps_1$ & 4.84E-06 &		-7.86E-06	&	-5.83E-06	&	-4.09E-05	&	9.98E-05 & 615 \\
 $\eps_2$ & 5.34E-06	&	-3.82E-06	&	-1.82E-05	&	-6.20E-06	&	-6.43E-05 & 70.1\\
$\eps_3$ & 5.62E-06	&	-5.68E-06	&	-2.01E-05	&	-9.23E-06	&	-3.44E-05 & 69.6\\
 $\eps_4$ & 1.66E-04 &		-4.30E-06&	-4.66E-05	&
 	-7.44E-05	&	-3.77E-05 & 29.5
  \\\hline
 $\eps_{C1}$ & -0.0015 &		-0.0017 & 	-0.0027 &		-0.0045 &	-0.0074	& 1,564\\
 $\eps_{C2}$ &-0.0024 &	-0.0029&	-0.00492&	-0.0084&	-0.013 & 546.3.3\\
 $\eps_{C3}$ &-0.010&	-0.0080&	-0.01&	-0.027&	-0.0103 & 865.3\\\hline
 $\eps_{Rich}$ &  -0.00058 &		-0.00050 &		-0.00043 &	 	-0.00063&	-0.0015  & 2,156.2
 \\\hline
 \end{tabular}
 \begin{flushleft}{\tiny
 Time: CPU time in msec, average of 100 runs\\
 BB: GWR-SINH(0;276,502): $r_0=0$, 276 points on $\cL^\pm$ in the main cycle and 502 points to evaluate the Wiener-Hopf 
 factors
 \\
 $\eps_0$: differences with GWR-SINH(0;306,557) with different omegas\\
$\eps_1,\eps_2, \eps_3,\eps_4$: differences with GWR-SINH(5;276,502),  GWR-SINH(0;108,188),  GWR-SINH(0.5;88,155), GWR-SINH(-0.5;55,100)\\
$\eps_{C1}, \eps_{C2}, \eps_{C3}$: errors of  Carr randomization method \cite{BLdouble} with $(N_T,\De_x)=(600,1/8000)$, 
$=(300,1/4000)$, $=(150,1/2000)$.
\\ $\eps_{Rich}$: errors of Richardson's extrapolation of C1 and C2 prices.
}
\end{flushleft}
\label{Table 3}
 \end{table}

\section{Pricing  barrier options in L\'evy models: general discussion and examples}\label{s:pricing_barr_Levy}
\subsection{A short review of the literature} The general formulas for single barrier options with continuous monitoring were derived in \cite{KoBoL,barrier-RLPE,NG-MBS}
using the operator form of the Wiener-Hopf factorziation \cite{eskin}, under certain regularity conditions on the characteristic exponent.
In \cite{single}, the same formulas were proved for any L\'evy process. 
The pricing formulas are in terms of Laplace-Fourier inversion in dimensions 2 (first touch digitals and no-touch options),
 3 (barrier puts and calls, and joint probability distributions of a L\'evy process and its extremum), and 4 (more general options with lookback and barrier features). Even marginally  accurate realizations of these formulas are far from trivial unless the characteristic exponent $\psi$ of the process is a rational function, hence, the Wiener-Hopf factors are rational functions as well. The factors are especially simple
 in the Double exponential jump-diffusion model (DEJD model) used in \cite{kou,KW1} and its generalization: Hyper-exponential jump-diffusion model (HEJD model) constructed independently in \cite{lipton-columbia,lipton-risk} (see also \cite{LiptonSelection}) and
\cite{amer-put-levy-maphysto,amer-put-levy}. 
In \cite{lipton-columbia,lipton-risk}, an explicit pricing formula for the joint distribution of
the L\'evy process and its extremum was derived using the Gaver-Stehfest algorithm (GS algorithm); the formula can be used to price options with barrier-lookback features. Later, a variation of the same technique was used in structural default models \cite{lipton-sepp}. In 
\cite{amer-put-levy-maphysto,amer-put-levy}, American options with finite time horizon are priced using the maturity randomization technique (Carr's randomization). The Wiener-Hopf factors in HEJD model are represented as weighted sums of exponential functions,
therefore, the calculations in the state space are reducible to application of convolution operators with exponentially decaying kernels.
A straightforward numerical realization of the convolution operators with exponential kernels is very fast: see \cite{amer-put-levy-maphysto,amer-put-levy,ExitRSw,stoch-int-rate-CF} for an explicit algorithm. An evident simplification of Carr's randomization can be applied to barrier options
(see \cite{MSdouble}, where double-barrier options in regime-switching models are priced): the early exercise boundary is fixed and it is unnecessary to fund
an approximation to the boundary at each step of backward induction. In both cases (GS-algorithm and Carr's randomization), the main block is the evaluation of
the perpetual options. If the GS-algorithm is used, it may be necessary to use high precision arithmetic because the weights are very large (see, e.g., examples
in \cite{paraLaplace}). The performance of the GS-algorithm can be improved using Wynn-Rho acceleration (GWR algorithm) - see \cite{AbateValko04} and,
 for examples in the context of pricing options of long maturity, \cite{paired}. If the GWR-algorithm can be used with double precision arithmetic, then, typically, the CPU time is smaller than
if Carr's randomization is applied. 
In the case of more general L\'evy processes, efficient calculations are much more difficult because the option price is very irregular at
the barrier and maturity. See Section \ref{ss:irr} for details and examples.  The irregular behavior makes it difficult to evaluate the prices of perpetual options sufficiently accurately so that the GWR-algorithm
or Carr's randomization used in \cite{paraLaplace,paired,UltraFast} can produce reasonably accurate results in regime-switchning models.  
The latter remark concerns other methods available in the literature - see, e.g.,  \cite{GSh,AAP,AKP,CV,beta,KudrLev09,KudrLev11,BIL,HaislipKaishev14,FusaiGermanoMarazzina,kirkbyJCompFinance18,Linetsky08,feng-linetsky09,LiLinetsky2015} and the bibliographies therein. The fundamental reasons for serious difficulties are as follows.

\subsection{Analysis of errors of different methods and an example}\label{ss:irr}

\begin{figure}
 \centering
    \begin{subfigure}[t]{0.45\textwidth}
{\includegraphics[width=\textwidth]{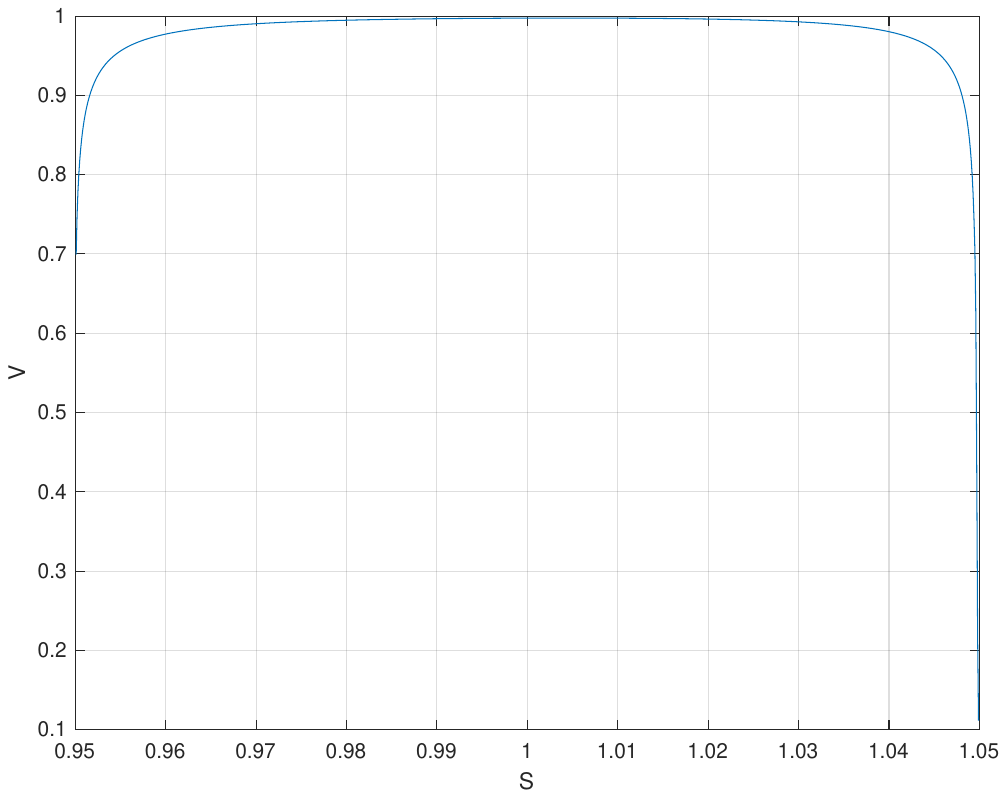}}\caption{$T=0.004$}
\end{subfigure}
\begin{subfigure}[t]{0.45\textwidth}
{\includegraphics[width=\textwidth]{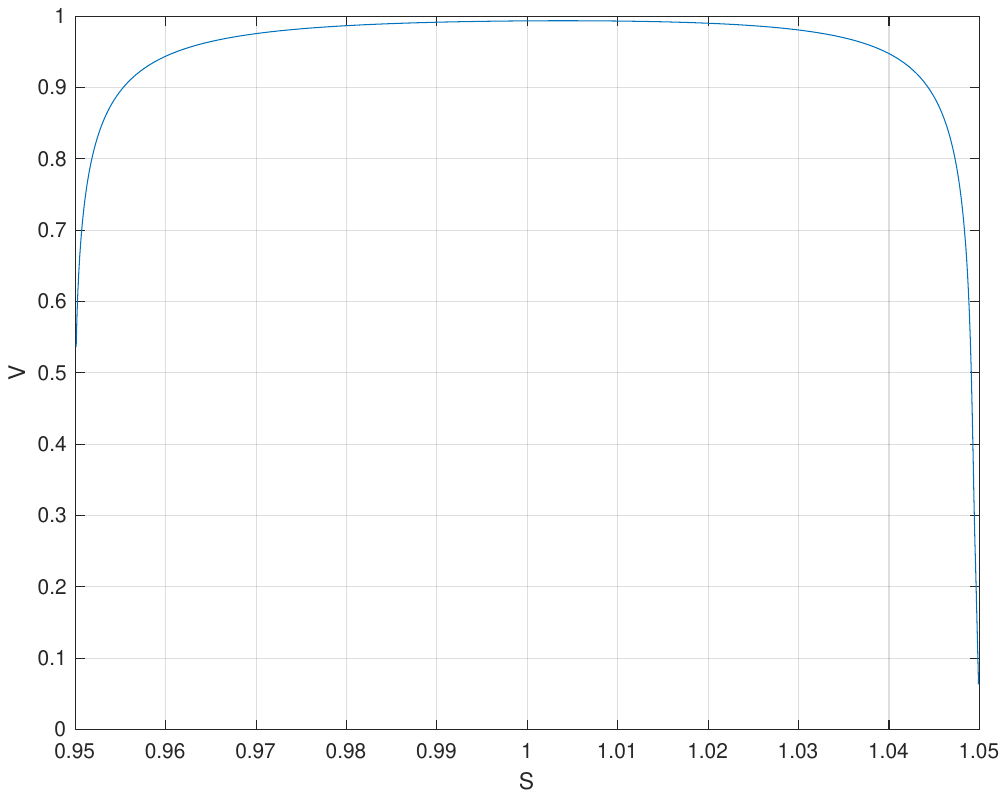}}\caption{$T=0.01$}
\end{subfigure}
 \caption{Prices of the double barrier option in KoBoL model. Parameters in Table \ref{Table 2} (AA). Barriers:
 $H_-=0.95, H_+=1.05$. }\label{fig:shortmaturity}

\end{figure}

The asymptotic analysis in \cite{NG-MBS,early-exercise,BIL,asymp-sens,AsAndersenLipton} demonstrates  that
if the characteristic exponent $\psi$ of a L\'evy process $X$ of any popular class is not a rational function, the price of a single barrier option is irregular at the boundary.\footnote{The asymptotic formulas are derived using a couple of general properties of $\psi$;
in \cite{EfficientAmenable} we verify these properties for all popular classes of processes.}
If $X$ is of finite variation and the drift points to (from) the boundary, then  the option price is of class $C^1$ up to the boundary
(is discontinuous at the boundary); option's gamma is unbounded in all cases. If $X$ is of infinite variation or the drift is 0, option's delta tend to infinity as the spot tends to the barrier. The representation of the Laplace transform of the double barrier options as a sum
of the Laplace transform of an European option and two perpetual single barrier options derived in \cite{MSdouble,BLdouble} and, in
the dual form, used in the present paper, implies that in all cases, the price of the double barrier option is irregular at one of the barriers,
at least. See Fig.~\ref{fig:shortmaturity} for an example.  In particular, in the vicinity of at least one barrier, the option price in a model with an irrational $\psi$
is larger than in a model with a rational characteristic exponent and in any model with non-negligible diffusion component.

If a numerical method uses the time discretization (and Carr's randomization can be interpreted as the time discretization),
and the ``true" process is replaced by a more convenient one so that the option price is of class $C^1$ up to the boundary,
then the errors in the vicinity of a boundary accumulate and propagate. If, at the same time, the number of time steps is large,
then the resulting error is quite large. See examples in \cite{single} which demonstrate sizable errors of
the approximation of KoBoL with HEJD: certainly, such an approximation will produces extremely large errors if applied
in regime-switching models. An example in Table \ref{Table 3} demonstrate, if Carr's randomization is used, it is difficult and time consuming to satisfy the moderate error tolerance of the order of 1\% unless the time step is very small. Fig.~\ref{fig:shortmaturity} clearly demonstrates
that if the time step is small, then the grid in the state space must be very fine, and the interpolation of order greater than 2 cannot be applied. Note the following important fact evident from Fig.~\ref{fig:shortmaturity} (the fact can be rigorously proved using 
the asymptotic analysis tools): the linear interpolation decreases the option price at each step of backward induction, which explains
why the Carr's randomization prices in Table \ref{Table 3} are smaller than the true prices. This effect dominates the well-known theoretical effect
of the time discretization: if, at each time step, the price is calculated perfectly or very accurately, the resulting price must be
larger than the true price. Depending on the ratio of the time step and step in the state space, either effect can dominate,
and one can choose a sequence of refinements of the grids in on the time line and state space which produce a sequence of results converging to any ``desirable" price - whether the latter is correct  or not.

Errors of approximations of the continuous time model with the discrete time model  are larger than the errors of Carr's randomization,
and accurate numerical realization of the transition operators for small time steps is very difficult because their kernels have a very high peak or even discontinuous at 0 (see \cite{NG-MBS}). Convenient approximations of the kernels using cosine functions (COS method) or B-spline approximations (BPROJ method) lead to serious errors. See \cite{MarcoDiscBarr} for examples of very large errors of COS method used to price
single barrier options, and discussion in \cite{BSINH} on the range of applicability of BPROJ method: since the error of the approximation is in $H^2$-norm, the approximation cannot work for options of very short maturity, for processes close to Variance Gamma especially.

Calculations in the dual space \cite{SINHregular,Contrarian,EfficientLevyExtremum} allows one to control errors efficiently,
and satisfy a small error tolerance using arrays of a moderate size. See Table \ref{Table 3}, where 
\begin{itemize}
\item
the number of nodes in the GWR algorithm is 16;
\item GWR-SINH($r_0$, $a$,$b$) means that  the grids on the curves $\cL^\pm$ in the dual space are of length $a$ in the main cycle, and length $b$ for the evaluation of the Wiener-Hopf factors;
\item
$N$ is the number of time steps in Carr's randomization algorithm, and $\De$ is the step of the grid in the $x=\ln S$ space.
\end{itemize}
Note that, in the presence of jumps, the lengths of the grids in the state space needed for accurate calculations are
larger than $\ln(H_+/H_-)/\De$. If the steepness parameters $\lp,\lm$ are not as large as in the examples (AA), (AB),
(MA), (MB) in the paper, then one has to use the grids several times longer and then the CPU time can be much larger than in the example shown in
Table \ref{Table 3}. The justification of the Richardson extrapolation for single barrier options in \cite{asymp-sens} admits a straightforward modification to the case of double barrier options\footnote{The Richardson extrapolation is not applicable to American options although widely used.} but as the last line in Table \ref{Table 3} shows, the improvement in the accuracy is not large.

 \section{Conclusion}\label{s:concl}
 In the paper, we constructed a new efficient method to price double barrier options in L\'evy models of finite variation and
non-zero drift. The method is a variation of the method in \cite{EfficientDoubleBarrier} for L\'evy processes of
either infinite variation or finite variation and zero drift. We explained why the method in \cite{EfficientDoubleBarrier}
 required a modification. In the operator language, the infinitesimal generator $L$ of a process in  \cite{EfficientDoubleBarrier} is a sectorial operator, and in the setting of the present paper $L$ is not sectorial, which makes it impossible to use efficient conformal deformations
of all contours of integration in the pricing formula. We use the Gaver-Wynn-Rho algorithm to evaluate the Bromwich integral,
and, for each value of the spectral parameter, calculate the price of the corresponding perpetual double barrier options making the calculations in the dual space. We calculate the resulting integrals using the sinh-acceleration technique \cite{SINHregular,Contrarian,EfficientLevyExtremum,EfficientDoubleBarrier}. The resulting GWR-SINH method is fast but less accurate
than the method in  \cite{EfficientDoubleBarrier} because the errors of the GWR algorithm for non-sectorial operators are larger. 
We discuss the sources of errors of popular groups of methods, which make it very difficult to price barrier options sufficiently accurately and fast so that arbitrage opportunity can be exploited, and present  empirical examples, where arbitrage opportunities may present themselves and efficient methods for pricing double barrier options are necessary.
A modification of the algorithm formulated in the paper can be used instead of the corresponding block in \cite{MSdouble} to price double barrier options in the regime-switching models and approximations of stochastic volatility models
and models with stochastic interest rates by regime-switching models that we developed earlier \cite{ExitRSw,stoch-int-rate-CF,amer-reg-sw-SIAM,SVolSSRN,BLHestonStIR08,BarrStIR}.


\appendix 



\section{Calibration results and prices of DNT options in the Heston and KoBoL models}\label{s: calib_and_DNT}
\begin{figure}
{\includegraphics[width=0.9\textwidth,height=0.5\textheight]{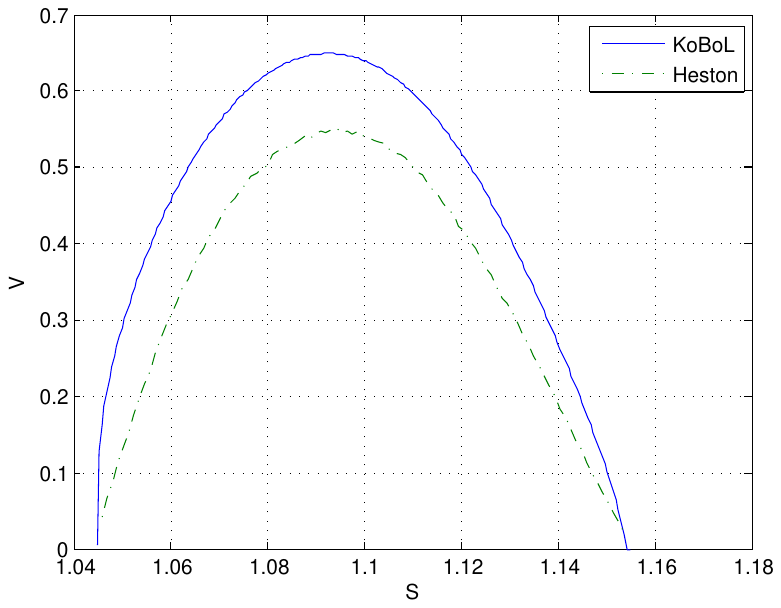}}
\caption{\small Prices of DNT under the Heston model and KoBoL. $T=0.25$, $S_0=1.09985, r_d=0.004, r_f=-0.01171$,
 $H_\pm=(1\pm 0.05)S_0$. The calibrated parameters are in Tables 1 and 2 (morning, ask).
 The irregularities of KoBoL prices near the barriers 
are due to insufficient accuracy of Carr's randomization algorithm used to calculate KoBoL prices.}
\label{fig:KoBoLandHestonAM}
\end{figure}

The empirical part of the present paper is the starting point of a project that we did several years ago. 
In \cite{SIAM2010}, a regime switching pure jump L\'evy model (KoBoL model modulated by a 2-3 state Markov chain) was suggested to 
account for jumps and stochastic volatility and skewness. Several years ago, we  applied the model to FX and studied the difference between the prices of double-no-touch options (DNT) in the Heston model, KoBoL model and 2-3 states KoBoL model
calibrated to the same real data in FX (Euro-USD pair; one day, the afternoon and morning bid and ask: AA, AB, MA, MB) using the standard procedure described in \cite{whystup_book_2e}. In Section \ref{s: calib_and_DNT}, we list the calibration results for the Heston 
and KoBoL models for all 4 cases (AA, AB, MA, MB) so that the reader can check that the vanilla prices in the Heston model and KoBoL model agree very well. Note that in all 4 cases, KoBoL of finite variation is documented, and the order of the process
(the analog of the Blumenthal-Getoor index for stable L\'evy processes) is close to 0.5. 
We observe that the prices of DNT options in KoBoL model are sizably larger than in the Heston model
(see Fig. \ref{fig:KoBoLandHestonAM}), as it is expected
by virtue of the asymptotic analysis of the behavior of the prices of single barrier options in \cite{NG-MBS,early-exercise,BIL,asymp-sens}.
A potentially important practical implication is that if the prices provided by Markit (in this particular situation, prices provided by Markit were fairly close
to the prices in the Heston model shown in Table \ref{DNTpricesHestonKoBoL})
 were the real quotes, 
one can use more accurate  purely jump models to exploit arbitrage opportunities. The discrepancy shown in 
Fig. \ref{fig:KoBoLandHestonAM} seem to be too large, and the calibration of a regime-switching 2-3-state KoBoL 
(hence, a multi-parameters model) gives  a host of candidates with similar goodness of fit characteristics; some of the candidates generate smaller discrepancies.
In the majority of cases, the prices are higher than the prices in the Heston model. 
To make a more definite statement, we lacked the real prices of DNT options, and the realization of
Carr's randomization method  \cite{BLdouble} that we used was too slow for accurate and fast calculations. See examples in Sect. \ref{s:GWR-SINH}.
In \cite{BLdouble}, the calculations are in the state space, which requires very long grids in the state space and dozens or even hundreds of time steps for options of short maturity to satisfy the error tolerance that the application to regime-switching models requires.

The results of calibration are collected in Tables \ref{Table 1}-\ref{Table 2}. 
We make the following remarks.

\begin{enumerate}
\item
There is no advantage to use a 5-parameter KoBoL. A quasi-symmetric 4-parameter version with
$c_+=c_-=c$ gives almost the best fit in all cases.\item
 Both a 5-parameter Heston model and a 4-parameter KoBoL fit the data
 very well. Various standard goodness of fit
 measures are fairly small and of the same order of magnitude for both models. In some cases,
 the goodness of fit of the 4-parameter KoBoL is better than that of the 5-parameter Heston.
 \item
 In all 4 cases, the goodness of fit (maximal relative error) of a 3-parameters KoBoL with fixed $\nu$ has
 a clear unique local minimum in the region $\nu\in (0.4, 0.6)$ (depending on the data).
 Hence, the underlying process is of infinite activity and finite variation.
 \item
 The dependence
 of other calibrated parameters on $\nu$ is fairly regular. Therefore,
 one can calibrate the model for several values of $\nu$
 using parallelization, interpolate the maximal relative error, find
 the minimizing $\nu$, and recalibrate the model
 either using this $\nu$ or do the search under the restriction that
 $\nu$ must be in a small neighborhood of the minimizer.
 The CPU time of calibration will decrease by a large factor.
 \item
 Goodness of fit to bids is sizably better than the one to asks. Probably, asks data is less stable
 because traders can easily afford to overshoot. 
 \item
 The calibrated parameters of the Heston model change more from the
 morning to afternoon than the KoBoL parameters, hence, the calibration of KoBoL is more stable.
 \item
 Parameters for bids differ significantly from the ones for ask
 but the prices of DNT  (see Table 4) are fairly close.
  \item
 In all 4 cases, the drift of KoBoL is positive which explains the sizable asymmetry in the tails.
 Since $r_d-r_f$ is positive and large, the results of calibration are not surprising.
  \item
 In the Heston case, the parameters change significantly from
 afternoon to morning (although the prices remain approximately the
 same). Hence, the volatility should be made stochastic. Similarly,
 for KoBoL, $c$ and $\nu$ should be made stochastic.  
 \item
 If, in the calibrated model, the drift is positive in one state, and
 negative in the other state (possibly, a two-state modulating Markov
 chain would be enough), then one expects that, in a neighborhood of
 the low barrier, the prices of DNT will be smaller, and in the
 neighborhood of the upper barrier, the behavior will be
 qualitatively similar to the one in the neighborhood of the lower
 barrier. Some asymmetry will remain, naturally, but one expect that
 the differences between the DNT prices under KoBoL and the ones
 under Heston in a neighborhood of $S_0$ will be of a similar order
 of magnitude. The limit of the price at each barrier will be
 positive but, likely, rather small.
 \item
 The difference between the DNT prices under the Heston model and the KoBoL model is quite sizable and systematic.
 The former prices are sizably smaller than the latter.


 \end{enumerate}

\begin{table}
\caption{\small Parameters of the Heston model (rounded), calibration to ask and bid prices, in the afternoon  (AA, AB), and in the  morning (MA, MB).
$\tau=0.25$, $r_d=0.004$, $r_f=-0.01171$; the afternoon spot $S_0=1.09865$, the morning spot $S_0=1.09985$}
{\tiny
\begin{tabular}{c|ccccc|cccc}
\hline
& $v_0$ & $\rho $ & $\sg_0$ & $\ka $ & $m$ &  rsme & ape & maxerr & maxrelerr\\\hline
AA & 0.00657 &	-0.2703 &	0.270 &	0.819 &	0.0283 &  8.78E-05	& 0.00176 &	0.000192 &	0.0104\\

AB & 0.00515 &	-0.364 &	0.250 &	0.943 &	0.0290 & 4.36E-06 &	9.74E-05 &	9.36E-06 &	0.000528\\\hline

MA &  0.00560 &	-0.267 &	0.274 &	0.152 &	0.178 &
     9.27E-05	& 0.00183 &	0.000201 &	0.0108\\

MB & 0.00441 &	-0.380 &	0.258	& 0.8367 &	0.0383 &
1.76E-05 &	0.000371 &	3.85E-05 &	0.00217\\\hline
\end{tabular}
}\label{Table 1}\end{table}

\begin{table}
\caption{\small Parameters of KoBoL (rounded), calibration to ask and bid prices, in the afternoon  (AA, AB), and in the  morning (MA, MB).
$T=0.25$, $r_d=0.004$, $r_f=-0.01171$; the afternoon spot $S_0=1.09865$, the morning spot $S_0=1.09985$. $\mu$ is the drift, $m_2$
is the second instantaneous moment.}
{\tiny
\begin{tabular}{c|cccccc|cccc}
\hline
& $c$ & $\nu$ & $\lp$ & $\lm $ & $\mu$ & $m_2$ &  rsme & ape & maxerr & maxrelerr\\\hline
AA & 0.881 &	0.491 &	25.71 &	-40.43 &	0.0718 &	0.00876 &
9.51E-05 &	0.00192 &	0.000208 &	0.0113\\
AB & 1.358 &	0.407 &	29.22 &	-52.14 &	0.09218 & 0.00784 & 2.94E-06 &	6.21E-05 &	6.43E-06	& 0.000363\\\hline

MA &  0.677 &	0.544	& 23.89 &	-37.69 &	0.0693 &	0.00894 &
     9.99E-05 &	0.0020 &0.000218 &	0.0118\\

MB & 1.125 &	0.445 &	27.93 &	-51.66 &	0.0940 &	0.00782 &
1.12E-05 &	0.00024	& 2.45E-05 &	0.00138\\\hline
\end{tabular}
}\label{Table 2}\end{table}

\begin{table}
\caption{\small Prices of the double no-touch options, in KoBoL and Heston models, calibrated to the afternoon and
the morning data, at the spot value $S_0$. The other parameters $T=0.25$, $r_d=0.004$, $r_f=-0.01171$.
 The barriers: $H_\pm=(1\pm 0.05)S_0$,  the afternoon spot $S_0=1.09865$, the morning spot $S_0=1.09985$.\\
 $V_H$: price under the Heston model calculated using the Milstein scheme with 500k paths and 10k times steps.\\
 $V_{KBL}$: price under the KoBoL model calculated using GWR-SINH method. \\
 $V_{KBL,MC}$: price under the KoBoL model calculated using the Monte-Carlo algorithm (Tankov's code http://www.math.jussieu.fr/~tankov/) with $\eps=E-09$, 2,000,000 trajectories and 20,000 time steps per trajectory.
 The standard deviation is in the range 0.012-0.018}

 \begin{tabular}{c|cc|cc}
\hline
 & AA & AB &  MA & MB \\
 \hline
 $V_{KBL}$ &  0.65266801 &	0.67764139 &0.65499963 & 0.68017579\\
 $V_{KBL,MC}$ & 0.6433 & 0.6675 & 0.6406 & 0.6702\\
 $Err_{MC}$ & -0.0101 & -0.094& -0.0144 & -0.0100\\
  $V_H$ &  0.5370 & 0.5718 & 0.5323 & 0.5742
    \\\hline
 \end{tabular}
 \label{DNTpricesHestonKoBoL}

 \end{table} 

 \section{Pricing barrier options in regime-switching models}\label{s:reg-switch}
 \subsection{The general scheme}
 For the sake of brevity, we consider a simplified version of the setting in  \cite{MSdouble}, with the same barriers in all states of the modulating Markov chain and no jumps in the underlying at switching moments.  
 Also, we restrict ourselves to the case of DNT options,
 with the constant terminal payoffs $G_j>0$ in all states of the modulating Markov chain; the reader can easily adjust the scheme to the case of more
 general payoff functions as in \cite{MSdouble}.
  
 Consider a regime-switching L\'evy model with finite state
modulating Markov chain denoted by $Y=(Y_t)$. The states are labelled by
$j=1,2,\ldots, m$, and the transition rate from state $j$ to state
$k\neq j$ by $\la_{jk}$.  The infinitesimal generator of $Y$ is
$L_Y=[\la_{jk}]$, where $\la_{jj}:=-\La_j$, and $\La_j=\sum_{k\neq
j}\la_{jk}$. For each $j$, we introduce a L\'evy process $X^{(j)}$
with the characteristic  exponent $\psi_j$ and infinitesimal
generator $L_j$, and
define
\[
X_t=\sum_{j=1}^m\int_0^t \bfo_{Y_u=j}\,dX^{(j)}_u.
\]
The processes $Y_t, X^{(1)}_t, \ldots, X^{(m)}_t$ are independent,
and jumps in $Y_t$ and $X^{(j)}_t$, $j=1,2,\ldots, m$, do not happen
simultaneously, a.s.
 Thus, we have the process
$\bX_t=(Y_t, X_t)$ with the state space $\{1,2,\ldots, m\}\times\bR$
and infinitesimal generator $\bL=L_Y\otimes I+{\rm diag}\, L_j$.
Under a risk-neutral measure $\bQ$ chosen for pricing, the underlying is modeled as $\log S_t=X_t$.

An additional factor $Y$ may account for stochastic volatility,
skewness or any other parameter that characterizes the dynamics of
the log-price while $Y$ remains in the same state.  We assume that
the riskless rate depends on the state, and is represented by an
$m$-tuple $\vec r=(r_j)_{j=1}^m$. Without loss of generality, we assume that $\vec r\ge 0$.

 The vector of the value functions $V(t,x)=[V_j(t,x)]_{j=1}^m$ in the states $j=1,2,\ldots,m$, at time $t<T$,
is the unique solution of the system
\beqast
V_j(t,x)&=&\bE^{x,t}\left[e^{-(T-t)(r_j+\La_j)}\bfo_{\tau^+_{h_+}\wedge \tau^-_{h_-}>T}G_j
\right.\\&&\left.
+\int_0^{\tau^+_{h_+}\wedge \tau^-_{h_-}\wedge T}e^{-(T-s)(r_j+\La_j)}\sum_{k\neq j}
\la_{jk}V_k(s,X^{(j)}_{s})dt\right],\ j=1,2,\ldots,m,
\eqast
 equivalently (see \cite{NG-MBS,barrier-RLPE} for general theorems about
 the correspondence between expectations of stochastic expressions and boundary problems), the unique bounded solution of the system
\beqast\label{rsw1}
(\dd_t+L_j-r_j-\La_j)V_j(t,x)&=&-\sum_{k\neq j}\la_{jk}V_k(t,x),\ t<T, x\in (h_-,h_+),\\\label{rsw2}
V_j(T,x)&=&G_j, \hskip3cm x\in (h_-,h_+),\\\label{rsw3}
V_j(t,x)&=&0, \hskip2.5cm t\le T,\ x\not\in (h_-,h_+).
\eqast
Introduce $Q_j(q):=q+\La_j+r_j$, $j=1,2,\ldots,m$, and  make the Laplace transform w.r.t. $\tau=T-t$.
 We obtain the system 
\beqa\label{rsqw12}
(Q_j(q)-L_j)\tV_j(q,x)&=&G_j+\sum_{k\neq j}\la_{jk}\tV_k(q,x),\ x\in (h_-,h_+),
\\\label{rsw32}
\tV_j(q,x)&=&0, \hskip2.5cm  x\not\in (h_-,h_+).
\eqa
For $q>0$, denote by $\tV^0(q)\in \bR^m$ the unique solution of the system
\bbe\label{V0j}
Q_j(q)\tV^{0}_j(q)= G_j+\sum_{k\neq j}\la_{jk} \tV^{0}_k(q),\ j=1,2,\ldots, m,
\ee
and set $\tV^{1}_{j}(q,x)=\tV_{j}(q,x)-V^{0}_j(q)$. The vector-function $\tV^{1}=[\tV^{1}_j]_{j=1}^m$
is a unique bounded solution of the system
\beqa\label{rsqw13}
(Q_j(q)-L_j)\tV^{1}_{j}(q,x)&=&\sum_{k\neq j}\la_{jk}\tV^{1}_{k}(q,x), \quad x\in (h_-,h_+),
\\\label{rsw33}
\tV^{1}_{j}(q,x)&=&-\tV^{0}_{j}(q), \hskip1.5cm \ x\not\in (h_-,h_+).
\eqa
 We calculate $\tV^1(q,x)$ in the form of the series 
similar to \eq{qtV}:
\beqa\label{qtVrsw}
 \tV^1(q,x)&=&\sum_{\ell =1}^{+\infty} (-1)^\ell (\tV^{+;\ell}(q,x)+\tV^{-;\ell}(q,x)).
 \eqa
  The terms of the series are vector functions as  $\tV^1(q,x)$ is. In the non-regime switching case,
 we have an explicit iteration procedure for calculation $\tV^{\pm;\ell}$ given $\tV^{\pm;\ell-1}$ (see
 \eq{def:tVp1}-\eq{def:tVmj}). In the regime-switching case, each of equations \eq{def:tVp1}-\eq{def:tVmj} becomes a system, which we solve using an embedded iteration procedure.
  Set $\tV^{-;0}(q,x)=\bfo_{[h_+,+\infty)}(x)\tV^0(q)$ and 
$\tV^{+;0}(q,x)=\bfo_{(-\infty,h_-]}(x)\tV^0(q)$, and, similarly to \eq{def:tVp1}, \eq{def:tVm1},
\eq{def:tVpj} and \eq{def:tVmj}, inductively define $\tV^{\pm;s}(q,x)=[\tV^{\pm;s}_{j}(q,x)]_{j=1}^m$, $s=1,2,\ldots,$ as the unique bounded
solutions of the systems
\beqa\label{rsqwp1}
(Q_j(q)-L_j)\tV^{+;s}_{j}(q,x)&=&\sum_{k\neq j}\la_{jk}\tV^{+;s}_{k}(q,x), \quad x<h_+,
\\\label{rswp2}
\tV^{+;s}_{j}(q,x)&=&\tV^{-;s-1}_{j}(q,x), \hskip1.2cm \ x\ge h_+,
\eqa
and
\beqa\label{rsqwm1}
(Q_j(q)-L_j)\tV^{-;s}_{j}(q,x)&=&\sum_{k\neq j}\la_{jk}\tV^{-;s}_{k}(q,x), \quad x>h_-,
\\\label{rswm2}
\tV^{-;s}_{j}(q,x)&=&\tV^{+;s-1}_{j}(q,x), \hskip1.2cm \ x\le h_-.
\eqa
Let $\cE_{Q_j(q)}, \cE^\pm_{Q_j(q)}$ be 
the EPV operators under $X^{(j)}$, the discount rate being $Q_j(q):=q+\La_j+r_j$. The general theorems for single barrier options in \cite{barrier-RLPE,NG-MBS}
(see also  \cite[Thm's 11.4.2-11.4.5]{IDUU})
allow us to rewrite the boundary problems \eq{rsqwp1}-\eq{rswp2} and \eq{rsqwm1}-\eq{rswm2} in the form
\beqa\label{rsqwp12}
\tV^{+;s}_{j}(q,x)&=&\frac{1}{Q_j(q)}\cE^+_{Q_j(q)}\bfo_{(-\infty,h_+)}\cE^-_{Q_j(q)}\sum_{k\neq j}\la_{jk}\tV^{+;s}_{k}(q,x)
\\\nonumber
&&+\cE^+_{Q_j(q)}\bfo_{[h_+,+\infty)}(\cE^+_{Q_j(q)})^{-1}\tV^{-;s-1}_{j}(q,x), \ j=1,2,\ldots, m,
 \eqa
 and
 \beqa\label{rsqwm12}
\tV^{-;s}_{j}(q,x)&=&\frac{1}{Q_j(q)}\cE^-_{Q_j(q)}\bfo_{(h_-,+\infty)}\cE^+_{Q_j(q)}\sum_{k\neq j}\la_{jk}\tV^{-;s}_{k}(q,x)
\\\nonumber
&&+\cE^-_{Q_j(q)}\bfo_{(-\infty,h_-]}(\cE^-_{Q_j(q)})^{-1}\tV^{+;s-1}_{j}(q,x), \ j=1,2,\ldots, m,
 \eqa
 respectively.   The systems \eq{rsqwp12} and \eq{rsqwm12} are solved using the straightforward modification
 of the iteration procedure in \cite{MSdouble}. Explicitly, let $s\ge 1$ be fixed and $\tV^{-;s-1}_{j}(q,\cdot)\in L_\infty(\bR)$ be given.
 Then the RHS' of the system \eq{rsqwp12} defines a contraction map from $L_\infty(\bR)$ to $L_\infty(\bR)$. Therefore,
 letting $\tV^{+;s;0}_{j}(q,\cdot)=0$ and, for $\ell=1,2,\ldots,$ defining 
 \beqa\label{rsqwp12iter}
\tV^{+;s;\ell}_{j}(q,x)&=&\frac{1}{Q_j(q)}\cE^+_{Q_j(q)}\bfo_{(-\infty,h_+)}\cE^-_{Q_j(q)}\sum_{k\neq j}\la_{jk}\tV^{+;s;\ell-1}_{k}(q,x)
\\\nonumber
&&+\cE^+_{Q_j(q)}\bfo_{[h_+,+\infty)}(\cE^+_{Q_j(q)})^{-1}\tV^{-;s-1}_{j}(q,x), \ j=1,2,\ldots, m,
 \eqa
 we conclude that $\tV^{+;s}_{j}(q,\cdot)=\lim_{\ell\to \infty}\tV^{+;s;\ell}_{j}(q,\cdot)$. The system \eq{rsqwm12} is solved similarly:
 we set $\tV^{-;s;0}_{j}(q,\cdot)=0$, then, for $\ell=1,2,\ldots,$ define
 \beqa\label{rsqwm12iter}
\tV^{-;s;\ell}_{j}(q,x)&=&\frac{1}{Q_j(q)}\cE^-_{Q_j(q)}\bfo_{(h_-,+\infty)}\cE^+_{Q_j(q)}\sum_{k\neq j}\la_{jk}\tV^{-;s;\ell-1}_{k}(q,x)
\\\nonumber
&&+\cE^-_{Q_j(q)}\bfo_{(-\infty,h_-]}(\cE^-_{Q_j(q)})^{-1}\tV^{+;s-1}_{j}(q,x), \ j=1,2,\ldots, m,
 \eqa
 and conclude that $\tV^{-;s}_{j}(q,\cdot)=\lim_{\ell\to \infty}\tV^{-;s;\ell}_{j}(q,x)$.
If the calculations are in the state space, then the same grids can and should be used for the numerical realization
of \eq{rsqwp12iter} and \eq{rsqwm12iter}, hence, one can use a straightforward variation of the algorithm  \cite{BLdouble} for the non-regime switching case. If the calculations are in the dual space (with the exception of the last step when 
the inverse Fourier transform is applied), then the basic algorithm of the non-regime switching case needs the following important modification. At each step of the iteration procedure, we need to evaluate the Fourier transforms 
$\htV^{+,s}_j(q,\xi)$ of functions $\tV^{+,s}_j(q,x)$ at {\em two non-intersecting curves in the lower half-plane}
and the Fourier transforms 
$\htV^{-,s}_j(q,\xi)$ of functions $\tV^{-,s}_j(q,x)$ at {\em two non-intersecting curves in the upper half-plane}.
 A similar trick (the Double Spiral method) is used in \cite{AsianGammaSIAMFM}  to price Asian options in L\'evy models.
 Note that one line can be used for all functions $\tV^{\pm,s}_j(q,x)$ if the Hilbert transform is used but
 the numerical realizations based on the Hilbert transform are less efficient than the sinh-acceleration. 
\subsection{Solution of the systems in the dual space}
With the exception of the last step, when the inverse Fourier transform is applied, the calculations are in the dual space.
To use
 the sinh-acceleration,
we have to use additional contours.
 The reason is the additional terms on the RHS' of \eq{rsqwp12} and \eq{rsqwm12}.
First, we choose two pairs of different straight lines $\cL^{0;\pm}_n=\{\Im\xi=\om_{n,\pm}\}$, $n=1,2,$ in the upper (sign $+$) and lower (sign $-$) half-plane; all are in the strip of analyticity of the Wiener-Hopf factors $\phi^\pm_q$. We may and shall assume that
$\om^-_2<\om^-_1<0<\om^+_1<\om^+_2$.  
 The Fourier transforms $\htV^+_j(q,\xi)$ of functions $\tV^+_j(q,x)$ are evaluated 
 on the curves $\cL^{0,-}_n;$ the Fourier transforms $\htV^-_j(q,\xi)$ of functions $\tV^-_j(q,x)$ are evaluated 
 on the curves $\cL^{0,+}_n,$ $n=1,2$. For $n=1,2$, 
 \beqa\label{htV0p}
 \htV^{+,0}(q,\xi)&=&\tV^0(q)\int_{h_+}^{+\infty}e^{-ix\xi}dx=\tV^0(q)\frac{e^{-ih_+\xi}}{i\xi}, \ \xi\in \cL^{0,-}_n,\\\label{htV0m}
 \htV^{,0}(q,\xi)&=&\tV^0(q)\int_{-\infty}^{h_-}e^{-ix\xi}dx=\tV^0(q)\frac{e^{-ih_-\xi}}{-i\xi}, \ \xi\in \cL^{0,+}_n,
 \eqa
and $\htV^{\pm;s;0}(q,\xi)=0$, $s\ge 1$. Thus, we have the basis of induction to evaluate the RHS of \eq{rsqwp12iter}
using the iteration procedure in the dual space,
for $s\ge 1$ and $\ell\ge 1$ fixed. Denote the  terms on the RHS of \eq{rsqwp12iter} by $\htV^{+;s;\ell}_{1,j}$ and $\htV^{+;s;\ell}_{2,j}$
so that 
\bbe\label{sumhtVj}
\htV^{+;s;\ell}_{j}=\htV^{+;s;\ell}_{1,j}+\htV^{+;s;\ell}_{2,j}.
\ee
The second term $\tV^{+;s;\ell}_{2,j}$ can be evaluated as in 
  the non-regime-switchnig case;
the only difference is that we evaluate the result for $\xi\in \cL^{0,-}_n$, $n=1,2$, and we can use either of the contours
$\cL^{0,+}_{n'}$, $n'=1,2$, for the integration:
  \beqast
\htV^{+;s;\ell}_{2,j}(q,\xi)&=&\phi^+_{Q_j(q)}(\xi)
\cF_{x\to \xi}\bfo_{[h_+,+\infty)}(\cE^+_q)^{-1}\tV^{-,s-1}_{j}(q,x)\\
&=&\phi^+_{Q_j(q)}(\xi)\int_{h_+}^{+\infty}dy\,e^{-iy\xi} \frac{1}{2\pi}\int_{\cL^{0,+}_{n'}}e^{i(y-h_-)\eta}
\phi^+_{Q_j(q)}(\eta)^{-1}\htV^{-;s-1}_{j}(q,\eta)d\eta.
 \eqast
 Since $\Im\xi<\om^+_{n'}$, the Fubini theorem is applicable and we can integrate w.r.t. $y$ first.
 The result is
 \bbe\label{rsqwp12iter1}
 \htV^{+;s;\ell}_{2,j}(q,\xi)=-\phi^+_{Q_j(q)}(\xi)e^{-ih_+\xi} \frac{1}{2\pi i}\int_{\cL^{0,+}_{n'}}\frac{e^{i(h_+-h_-)\eta}}{\eta-\xi}
 \phi^+_{Q_j(q)}(\eta)^{-1}\htV^{-;s-1}_{j}(q,\eta)d\eta.
 \ee
 The evaluation of $\htV^{+;s;\ell}_{1,j}$ on $\cL^{0,-}_n$, $n=1,2$, is more involved because the terms $\htV^{+;s;\ell}_{k}$ are calculated at the previous induction step on the same pair of contours.  In order to avoid the  singularity at $\eta=\xi$, we
 integrate over $\cL^{0,-}_1$ to evaluate $\htV^{+;s;\ell}_{1,j}$ on $\cL^{0,-}_2$, and over $\cL^{0,-}_2$ to evaluate $\htV^{+;s;\ell}_{1,j}$ on $\cL^{0,-}_1$.
 Since $\cL^{0,-}_1$ is a horizontal line above $\cL^{0,-}_2$, essentially the same argument as in the proof of \eq{rsqwp12iter1}
 gives: for $\xi\in \cL^{0,-}_1$,
 \bbe\label{rsqwp12iter21}
 \htV^{+;s;\ell}_{1,j}(q,\xi)= \frac{1}{Q_j(q)}\phi^+_{Q_j(q)}(\xi)\frac{e^{-ih_+\xi} }{2\pi i}\int_{\cL^{0,-}_{2}}\frac{e^{i(h_+-h_-)\eta}}{\eta-\xi}
\phi^-_{Q_j(q)}(\eta)\sum_{k\neq j}\la_{jk}\htV^{+;s;\ell-1}_{j}(q,\eta)d\eta.
 \ee
 To obtain the representation for $ \htV^{+;s;\ell}_{1,j}(q,\xi), \xi\in \cL^{0,-}_2$, with the integration over $\cL^{0,-}_1$,
 we start with \eq{rsqwp12iter21} for $\xi$ in the open strip sandwiched between the two contours, push the line of integration up and, on crossing the singularity at $\eta=\xi$, apply the residue theorem
 and the Wiener-Hopf identity:
  \beqa\label{rsqwp12iter22}
 \htV^{+;s;\ell}_{1,j}(q,\xi)&=&\frac{\phi^+_{Q_j(q)}(\xi)}{Q_j(q)} \frac{e^{-ih_+\xi}}{2\pi i}\int_{\cL^{0,-}_{1}}\frac{e^{i(h_+-h_-)\eta}}{\eta-\xi}
 \phi^-_{Q_j(q)}(\eta)\sum_{k\neq j}\la_{jk}\htV^{+;s;\ell-1}_{j}(q,\eta)d\eta\\\nonumber
 &&+(Q_j(q)+\psi^{j}(\xi))^{-1}\sum_{k\neq j}\la_{jk}\htV^{+;s;\ell-1}_{j}(q,\xi).
 \eqa
 Finally, we notice that, by the uniqueness of analytic continuation,
 the result is valid for $\xi\in \cL^{0,-}_2$.
 The set of formulas \eq{sumhtVj}, \eq{rsqwp12iter1} and \eq{rsqwp12iter21} for $\xi\in \cL^{0,-}_1$, \eq{rsqwp12iter22} for $\xi\in \cL^{0,-}_2$, define a system of integral equations for the restrictions of  $\htV^{+;s;\ell}_{j}(q,\xi)$ on the union
 $\cL^{0,-}_1\cup\cL^{0,-}_2$. Using the sinh-acceleration technique, we can deform both contours $\cL^{0,-}_n$, $n=1,2$, and, simultaneously,
 the contours $\cL^{0,+}_{n'}, n'=1,2,$ in the formulas \eq{rsqwp12iter1}, \eq{rsqwp12iter21}, \eq{rsqwp12iter22}.
 In the formulas for the sinh-deformation $\xi^\pm_n=i\om^\pm_1+b^\pm_n\sinh(i\om^\pm_n+y)$, we choose the parameters of the deformed curves, denote them $\cL^\pm_n, n=1,2,$
 so that the curves do not intersect and, in the new coordinates, the strips of analyticity are wide.
 In the case of SL-processes, a natural recommendation is to take $\om^\pm_1=\pm \pi/6, \om^\pm_2=\pm\pi/3$.
 The other parameters are chosen so that the curves do not intersect and are located in the upper and lower half-planes.
 
 The system \eq{rsqwm12iter} is reduced to the system of integral equations by symmetry: the value functions $\htV^{-;s;\ell}$ are defined on the lines $\cL^{0,+}_n, n=1,2,$, the integration is over either of the lines $\cL^{0,-}_{n'}, n'=1,2$,
 and the same contour deformations are used.

\end{document}